\documentclass[twocolumn, trackchanges]{aastex62}

\submitjournal{ApJ}
\usepackage{amsmath}
\usepackage{comment}
\usepackage{cleveref}

\usepackage[utf8]{inputenc}

\crefformat{section}{\S#2#1#3} 
\crefformat{subsection}{\S#2#1#3}
\crefformat{subsubsection}{\S#2#1#3}

\shortauthors{Rinaldi et al.}

\begin{document}

\title{\bf Deciphering the Nature of {\it Virgil}: An Obscured AGN Lurking Within an Apparently Normal Lyman-$\alpha$ Emitter During Cosmic Reionization}

\newcommand{\gsim}{{\;\raise0.3ex\hbox{$>$\kern-0.75em\raise-1.1ex\hbox{$\sim$}}\;}}

\correspondingauthor{Pierluigi Rinaldi}
\email{prinaldi@arizona.edu}

\author[0000-0002-5104-8245]{Pierluigi Rinaldi}
\affiliation{Steward Observatory, University of Arizona, 933 North Cherry Avenue, Tucson, AZ 85721, USA}

\author[0000-0003-4528-5639]{Pablo G. P\'erez-Gonz\'alez}
\affiliation{Centro de Astrobiolog\'ia (CAB), CSIC–INTA, Cra. de Ajalvir Km.~4, 28850- Torrej\'on de Ardoz, Madrid, Spain}

\author[0000-0003-2303-6519]{George H. Rieke}
\affiliation{Steward Observatory, University of Arizona, 933 North Cherry Avenue, Tucson, AZ 85721, USA}

\author[0000-0002-6221-1829]{Jianwei Lyu}
\affiliation{Steward Observatory, University of Arizona, 933 North Cherry Avenue, Tucson, AZ 85721, USA}

\author[0000-0003-2388-8172]{Francesco D'Eugenio}
\affiliation{Kavli Institute for Cosmology, University of Cambridge, Madingley Road, Cambridge, CB3 0HA, UK}
\affiliation{Cavendish Laboratory, University of Cambridge, 19 JJ Thomson Avenue, Cambridge, CB3 0HE, UK}

\author[0000-0002-8876-5248]{Zihao Wu}
\affiliation{Center for Astrophysics $|$ Harvard \& Smithsonian, 60 Garden St., Cambridge MA 02138 USA}

\author[0000-0002-6719-380X]{Stefano Carniani}
\affiliation{Scuola Normale Superiore, Piazza dei Cavalieri 7, I-56126 Pisa, Italy}

\author[0000-0002-3642-2446]{Tobias J. Looser}
\affiliation{Kavli Institute for Cosmology, University of Cambridge, Madingley Road, Cambridge, CB3 0HA, UK}
\affiliation{Cavendish Laboratory, University of Cambridge, 19 JJ Thomson Avenue, Cambridge, CB3 0HE, UK}

\author[0000-0003-4702-7561]{Irene Shivaei}
\affiliation{Centro de Astrobiolog\'ia (CAB), CSIC–INTA, Cra. de Ajalvir Km.~4, 28850- Torrej\'on de Ardoz, Madrid, Spain}

\author[0000-0002-3952-8588]{Leindert A. Boogaard}
\affiliation{Leiden Observatory, Leiden University, PO Box 9513, NL-2300 RA Leiden, The Netherlands}

\author[0000-0003-0699-6083]{Tanio Diaz-Santos}
\affiliation{Institute of Astrophysics, Foundation for Research and Technology-Hellas (FORTH), Heraklion, 70013, Greece}
\affiliation{School of Sciences, European University Cyprus, Diogenes street, Engomi, 1516 Nicosia, Cyprus}

\author[0000-0002-9090-4227]{Luis Colina}
\affiliation{Centro de Astrobiolog\'ia (CAB), CSIC–INTA, Cra. de Ajalvir Km.~4, 28850- Torrej\'on de Ardoz, Madrid, Spain}

\author[0000-0002-3005-1349]{G\"oran \"Ostlin}
\affiliation{Department of Astronomy, Stockholm University, Oscar Klein Centre, AlbaNova University Centre, 106 91 Stockholm, Sweden}

\author[0000-0002-8909-8782]{Stacey Alberts}
\affiliation{Steward Observatory, University of Arizona, 933 North Cherry Avenue, Tucson, AZ 85721, USA}

\author[0000-0002-7093-1877]{Javier \'Alvarez-M\'arquez}
\affiliation{Centro de Astrobiolog\'ia (CAB), CSIC–INTA, Cra. de Ajalvir Km.~4, 28850- Torrej\'on de Ardoz, Madrid, Spain}

\author[0000-0002-8053-8040]{Marianna Annuziatella}
\affiliation{Centro de Astrobiolog\'ia (CAB), CSIC--INTA, Cra. de Ajalvir Km.~4, 28850- Torrej\'on de Ardoz, Madrid, Spain}

\author[0000-0002-6290-3198]{Manuel Aravena}
\affiliation{Instituto de Estudios Astrof\'isicos, Facultad de Ingenier\'ia y Ciencias, Universidad Diego Portales, Av. Ej\'ercito 441, Santiago 8370191, Chile}

\author[0000-0003-0883-2226]{Rachana Bhatawdekar}
\affiliation{European Space Agency (ESA), European Space Astronomy Centre (ESAC), Camino Bajo del Castillo s/n, 28692 Villanueva de la Ca\~nada, Madrid, Spain}

\author[0000-0002-8651-9879]{Andrew J. Bunker}
\affiliation{Department of Physics, University of Oxford, Denys Wilkinson Building, Keble Road, Oxford OX1 3RH, UK}

\author[0000-0001-8183-1460]{Karina I. Caputi}
\affiliation{Kapteyn Astronomical Institute, University of Groningen,
P.O. Box 800, 9700AV Groningen,
The Netherlands}
\affiliation{Cosmic Dawn Center (DAWN), Copenhagen, Denmark}

\author[0000-0003-3458-2275]{St\'ephane Charlot}
\affiliation{Sorbonne Universit\'e, CNRS, UMR 7095, Institut d'Astrophysique de Paris, 98 bis bd Arago, 75014 Paris, France}

\author[0000-0003-2119-277X]{Alejandro Crespo G\'omez}
\affiliation{Space Telescope Science Institute (STScI), 3700 San Martin Drive, Baltimore, MD 21218, USA}

\author[0000-0002-2678-2560]{Mirko Curti}
\affiliation{European Southern Observatory, Karl-Schwarzschild-Strasse 2, 85748 Garching, Germany}

\author[0000-0000-0000-0000]{Andreas Eckart}
\affiliation{Physikalisches Institut der Universität zu Köln, Zülpicher Str. 77, 50937 Köln, Germany}

\author[0000-0001-9885-4589]{Steven Gillman}
\affiliation{Cosmic Dawn Center (DAWN), Copenhagen, Denmark}
\affiliation{DTU Space, Technical University of Denmark, Elektrovej, Building 328, 2800, Kgs. Lyngby, Denmark}

\author[0000-0003-4565-8239]{Kevin Hainline}
\affiliation{Steward Observatory, University of Arizona, 933 North Cherry Avenue, Tucson, AZ 85721, USA}

\author[0000-0002-5320-2568]{Nimisha Kumari}
\affiliation{AURA for European Space Agency, Space Telescope Science Institute, 3700 San Martin Drive. Baltimore, MD, 21210}

\author[0000-0002-4571-2306]{Jens Hjorth}
\affiliation{DARK, Niels Bohr Institute, University of Copenhagen, Jagtvej 155A, 2200 Copenhagen, Denmark}

\author[0000-0001-8386-3546]{Edoardo Iani}
\affiliation{Institute of Science and Technology Austria (ISTA), Am Campus 1, 3400 Klosterneuburg, Austria}

\author[0000-0003-4268-0393]{Hanae Inami}
\affiliation{Hiroshima Astrophysical Science Center, Hiroshima University, 1-3-1 Kagamiyama, Higashi-Hiroshima, Hiroshima 739-8526, Japan}

\author[0000-0001-7673-2257]{Zhiyuan Ji}
\affiliation{Steward Observatory, University of Arizona, 933 North Cherry Avenue, Tucson, AZ 85721, USA}

\author[0000-0002-9280-7594]{Benjamin D.\ Johnson}
\affiliation{Center for Astrophysics $|$ Harvard \& Smithsonian, 60 Garden St., Cambridge MA 02138 USA}

\author[0000-0002-4201-7367]{Gareth C.\ Jones}
\affiliation{Kavli Institute for Cosmology, University of Cambridge, Madingley Road, Cambridge, CB3 0HA, UK}
\affiliation{Cavendish Laboratory, University of Cambridge, 19 JJ Thomson Avenue, Cambridge, CB3 0HE, UK}

\author[0000-0002-0690-8824]{\'Alvaro Labiano}
\affiliation{Telespazio UK for the European Space Agency, ESAC, Camino Bajo del Castillo s/n, 28692 Villanueva de la Ca\~nada, Spain}

\author[0000-0002-4985-3819]{Roberto Maiolino}
\affiliation{Kavli Institute for Cosmology, University of Cambridge, Madingley Road, Cambridge, CB3 0HA, UK}
\affiliation{Cavendish Laboratory, University of Cambridge, 19 JJ Thomson Avenue, Cambridge, CB3 0HE, UK}
\affiliation{Department of Physics and Astronomy, University College London, Gower Street, London WC1E 6BT, UK}

\author[0000-0003-0470-8754]{Jens Melinder}
\affiliation{Department of Astronomy, Stockholm University, Oscar Klein Centre, AlbaNova University Centre, 106 91 Stockholm, Sweden}

\author[0000-0002-3305-9901]{Thibaud Moutard}
\affiliation{European Space Agency (ESA), European Space Astronomy Centre (ESAC), Camino Bajo del Castillo s/n, 28692 Villanueva de la Ca\~nada, Madrid, Spain}
\affiliation{Aix Marseille Univ, CNRS, CNES, LAM, Marseille, France}

\author[0000-0000-0000-0000]{Florian Pei\ss ker} 
\affiliation{Physikalisches Institut der Universität zu Köln, Zülpicher Str. 77, 50937 Köln, Germany}

\author[0000-0002-7893-6170]{Marcia Rieke}
\affiliation{Steward Observatory, University of Arizona, 933 North Cherry Avenue, Tucson, AZ 85721, USA}

\author[0000-0002-4271-0364]{Brant Robertson}
\affiliation{Department of Astronomy and Astrophysics, University of California, Santa Cruz, 1156 High Street, Santa Cruz, CA 95064, USA}

\author[0000-0001-6010-6809]{Jan Scholtz}
\affiliation{Kavli Institute for Cosmology, University of Cambridge, Madingley Road, Cambridge, CB3 0HA, UK}
\affiliation{Cavendish Laboratory, University of Cambridge, 19 JJ Thomson Avenue, Cambridge, CB3 0HE, UK}

\author[0000-0002-8224-4505]{Sandro Tacchella}
\affiliation{Kavli Institute for Cosmology, University of Cambridge, Madingley Road, Cambridge, CB3 0HA, UK}
\affiliation{Cavendish Laboratory, University of Cambridge, 19 JJ Thomson Avenue, Cambridge, CB3 0HE, UK}

\author[0000-0001-5434-5942]{Paul P. van der Werf}
\affiliation{Leiden Observatory, Leiden University, PO Box 9513, NL-2300 RA Leiden, The Netherlands}

\author[0000-0003-4793-7880]{Fabian Walter}
\affiliation{Max Planck Institut f\"ur Astronomie, K\"onigstuhl 17, D-69117, Heidelberg, Germany}

\author[0000-0003-2919-7495]{Christina C. Williams}
\affiliation{NSF National Optical-Infrared Astronomy Research Laboratory, 950 North Cherry Avenue, Tucson, AZ 85719, USA}

\author[0000-0002-4201-7367]{Chris Willott}
\affiliation{NRC Herzberg, 5071 West Saanich Rd, Victoria, BC V9E 2E7, Canada}

\author[0000-0002-7595-121X]{Joris Witstok}
\affiliation{Cosmic Dawn Center (DAWN), Copenhagen, Denmark}
\affiliation{Niels Bohr Institute, University of Copenhagen, Jagtvej 128, DK-2200, Copenhagen, Denmark}

\author[0000-0003-4891-0794]{Hannah \"Ubler}
\affiliation{Max-Planck-Institut f\"ur extraterrestrische Physik (MPE), Gie\ss enbachstra\ss e 1, 85748 Garching, Germany}

\author[0000-0003-3307-7525]{Yongda Zhu}
\affiliation{Steward Observatory, University of Arizona, 933 North Cherry Avenue, Tucson, AZ 85721, USA}

\begin{abstract}

We present a comprehensive analysis of the MIRI Extremely Red Object \textit{Virgil}, a Lyman-$\alpha$ emitter at $z_{spec} = 6.6379 \pm0.0035$ with the photometric properties of a Little Red Dot. Leveraging new JWST/MIRI imaging from the MIDIS and PAHSPECS programs, we confirm \textit{Virgil}'s extraordinary nature among galaxies in JADES/GOODS-South, exhibiting a strikingly red NIRCam-to-MIRI color (F444W $-$ F1500W = $2.84\pm0.04$~mag). Deep NIRSpec/PRISM spectroscopy from the OASIS program offers key insights into the host galaxy, revealing properties of an average star-forming galaxy during Cosmic Reionization, such as a subsolar metallicity, low-to-moderate dust content, and a relatively high ionization parameter and electron temperature. By estimating the star formation rate of \textit{Virgil} from UV and H$\alpha$, we find evidence that the galaxy is either entering or fading out of a bursty episode. Although line-ratio diagnostics employed at high-$z$ would classify \textit{Virgil} as an Active Galactic Nucleus (AGN), this classification becomes ambiguous once redshift evolution is considered. Nonetheless, \textit{Virgil} occupies the same parameter space as recently confirmed AGNs at similar redshifts. The new deep MIRI data at 15~$\mu$m reinforce the AGN nature of \textit{Virgil}, as inferred from multiple spectral energy distribution (SED) fitting codes. \textit{Virgil}'s rising infrared SED and UV excess resemble those of Dust-Obscured Galaxies (DOGs) studied with {\it Spitzer} at Cosmic Noon, particularly blue-excess HotDOGs. Our results highlight the need for a multi-wavelength approach incorporating MIRI to uncover such extreme sources at $z\gtrsim6$ and to shed light on the interplay between galaxy evolution and early black hole growth during Cosmic Reionization.
\end{abstract}


\section{Introduction}

Infrared (IR) astronomy has advanced dramatically in recent decades. The Infrared Astronomical Satellite (IRAS; \citealt{neugebauer_infrared_1984}) and the Infrared Space Observatory (ISO; \citealt{helou_infrared_1995}) provided the first all-sky IR surveys, laying the foundation for extragalactic studies. These efforts paved the way for the \textit{Spitzer} Space Telescope \citep{gehrz_nasa_2007}—\textit{the last of the Great Observatories}\footnote{As described in  \citet{rieke_last_2006}.}—and later, the \textit{Herschel} Space Observatory \citep{pilbratt_Herschel_2010}. While \textit{Herschel} extended coverage into the far-IR, \textit{Spitzer} remained unmatched in mid-IR sensitivity and resolution throughout the pre-\textit{JWST} era \citep{gardner_james_2023}.

With the Infrared Array Camera (IRAC; \citealt{fazio_infrared_2004}) and the Multiband Imaging Photometer (MIPS; \citealt{rieke_multiband_2004}), \textit{Spitzer} allowed us to advance our understanding of galaxy evolution. It enabled robust estimates of stellar mass ($M_\star$), age, and star formation rate (SFR) in high-redshift galaxies \citep[e.g.,][]{papovich_spitzer_2006, yan_stellar_2006, bradac_high-redshift_2020}, uncovered massive systems at $z > 6$, and observed some of the most distant sources known at the time—e.g., GN-z11 \citep{oesch_most_2014, oesch_remarkably_2016} and MACS0647-JD \citep{lam_detection_2019, strait_stellar_2020}. It also revealed strong nebular emission through photometric excesses \citep[e.g.,][]{huang_spitzer_2016}, Balmer breaks, and early dust content \citep{papovich_spitzer_2006}. In parallel, it advanced studies of active galactic nuclei (AGNs; \citealt{lacy_active_2020}) and clarified the role of luminous and ultraluminous infrared galaxies (LIRGs/ULIRGs) in driving feedback between starbursts, black holes, and the interstellar medium (ISM; \citealt{armus_observations_2020}). Many of these phenomena—such as early dust and strong emission lines—are now confirmed by \textit{JWST} photometry and spectroscopy \citep{rinaldi_midis_2023, boyett_extreme_2024, kuruvanthodi_strong_2024, langeroodi_rapid_2024}.

In particular, \textit{Spitzer} played a key role in advancing the study of Extremely Red Objects (EROs; \citealt{elston_deep_1988, elston_observations_1989})—sources with very red mid-IR to optical colors—triggering extensive investigations into their nature \citep[e.g.,][]{wilson_extremely_2004, yan_high-redshift_2004, stern_spitzer_2006, wilson_aegis_2007}. Initially believed to be $z \approx 10$ sources \citep[e.g.,][]{mobasher_evidence_2005}, EROs were later found to include Submillimeter Galaxies (SMGs; \citealt{blain_submillimeter_2002}) and Dust-Obscured Galaxies (DOGs; \citealt{dey_significant_2008}). DOGs—categorized as power-law or “bump” types (with a SED flattening at long wavelengths) —likely trace a transition from dusty star formation to AGN activity during gas-rich mergers, with their red colors driven by heavy obscuration and mid-IR re-emission.  Additionally, EROs were found to overlap with high-redshift galaxies (\citealt{bouwens_ultraviolet_2011}) selected via the Lyman break technique \citep{steidel_spectroscopy_1996}, highlighting their heterogeneous nature.

Despite its breakthroughs, \textit{Spitzer} was limited by its spatial resolution. This limitation has now been
overcome by JWST, which combines unparalleled sensitivity, angular resolution, and wavelength coverage with both the Near-Infrared Camera (NIRCam, 0.7--4.8 $\mu$m; \citealt{rieke_performance_2023}) and the Mid-Infrared Instrument (MIRI, 5.6--25.5\footnote{Here we refer to the MIRI imager. However, the MIRI Medium Resolution Spectroscopy (MRS) extends up to $\approx$ 27 $\mu$m.} $\mu$m; \citealt{wright_mid-infrared_2023}). The launch of JWST marks a significant leap forward compared to its predecessors, including the {\it Hubble} Space Telescope (HST), heralding a {\it golden era} of IR extragalactic astronomy.

Since its launch, JWST has pushed the boundaries of the redshift frontier, with groundbreaking discoveries at very high redshift, such as GS-z12-0, GS-z13-0, GS-z14-0, and GS-z14-1 \citep{curtis-lake_spectroscopic_2023, carniani_shining_2024, bunker_jades_2024} from the JWST Advanced Deep Extragalactic Survey (JADES; \citealt{bunker_spectroscopy_2020, eisenstein_jades_2023, rieke_jades_2023}) team, as well as other very high-$z$ sources identified by independent teams \citep[e.g.,][]{arrabal_haro_confirmation_2023, wang_uncover_2023}. These observations have offered unprecedented insights into the very early Universe, with the Near Infrared Spectrograph’s Micro-Shutter Assembly (NIRSpec/MSA; \citealt{ferruit_near-infrared_2022, jakobsen_near-infrared_2022}) playing a crucial role in spectroscopically confirming galaxies at $z > 10$.

With JWST, the ERO “industry” is back in the spotlight, unveiling even more compact and red sources (\citealt{labbe_population_2023}), now known as “Little Red Dots” (LRDs; \citealt{matthee_little_2024}). The study of these sources and their nature has triggered a huge amount of literature in a very short time (e.g., \citealt{furtak_jwst_2023, killi_deciphering_2023, kokorev_uncover_2023, ubler_ga-nifs_2023, akins_cosmos-web_2024, barro_extremely_2024, greene_uncover_2024, kocevski_rise_2024, kokorev_census_2024, hainline_investigation_2024, perez-gonzalez_what_2024, rinaldi_not_2024, williams_galaxies_2024}).

Recent studies highlighted MIRI’s critical role in uncovering EROs, revealing “exotic” objects undetectable even in the deepest NIRCam surveys, such as JADES \citep{eisenstein_jades_2023, rieke_jades_2023}, or showing unexpected spectral energy distribution (SED) upturns in objects that NIRCam would typically classify as star-forming galaxies (SFGs). In this context, the MIRI Deep Imaging Survey (MIDIS; \citealt{ostlin_miri_2024}) has been fundamental in identifying the first MIRI Extremely Red Objects (MEROs): {\it Cerberus} \citep{perez-gonzalez_nircam-dark_2024} and {\it Virgil} \citep{iani_midis_2024}.

While the nature of \textit{Cerberus}—a source uniquely detected at 10~$\mu$m—remains entirely unknown because it is so faint, \textit{Virgil} offers a great opportunity for detailed investigation thanks to its detection across multiple facilities, including HST, JWST (NIRCam and MIRI), and VLT/MUSE \citep{bacon_muse_2023}. As reported in \citet{iani_midis_2024}, {\it Virgil} is identified as a Lyman-$\alpha$ emitter (LAE) at $z_{\mathrm{spec}} = 6.6312 \pm 0.0019$ (from VLT/MUSE) located in GOODS-South (GOODS-S; \citealt{dickinson_great_2003, giavalisco_great_2004}). It exhibits very red colors between NIRCam and MIRI bands (F444W $-$ F1000W $> 2$~mag; \citealt{iani_midis_2024}), thus revealing that the MIRI photometric information is crucial at $z\gtrsim6$ to identify it and similar objects, as the strong upturn happens entirely outside the wavelength range of NIRCam.

\citet{iani_midis_2024} concluded that \textit{Virgil} could be either a dusty starburst—similar to what has been proposed for GN20 (e.g., \citealt{colina_uncovering_2023})—or a SFG hosting a dust-obscured AGN. Recent JWST/NIRSpec observations of GN20 reveal complex kinematics, extended H$\alpha$ emission, and signatures of AGN-driven feedback \citep{ubler_ga-nifs-gn20_2024}. However, unlike GN20, \textit{Virgil} is both much less massive and more compact, making its extreme red colors and potential AGN activity even more remarkable.
In this case, no AGN counterparts were identified in any of the existing multi-wavelength catalogs of GOODS-S (\citealt{ranalli_xmm_2013, luo_chandra_2017, lyu_agn_2022, gillman_midis_2025}), including the most recent AGN catalog from \citet{lyu_active_2024} with MIRI data, which is based on the Systematic Mid-infrared Instrument Legacy Extragalactic Survey (SMILES; \citealt{alberts_smiles_2024, rieke_smiles_2024}). In the latter case, this could be due to the limited depth of the SMILES observations, which are not deep enough to clearly detect ($>3\sigma$) this object in the MIRI bands above 7.7 $\mu$m. 

Moreover, \citet{iani_midis_2024} demonstrated that {\it Virgil}'s SED closely resembles that of a typical LRD (see their Figure 10). Although {\it Virgil} does not meet all photometric color criteria for LRDs (e.g., \citealt{kokorev_census_2024}), it satisfies the compactness criterion in the F444W band, possibly making it one of the first LRDs where the host galaxy is well-detected. Interestingly, \citet{iani_midis_2024} highlight the peculiar ultraviolet (UV) morphology of {\it Virgil}, consistent with the findings of \citet{rinaldi_not_2024}, where nearly 30\% of the photometrically selected LRDs exhibit a complex UV morphology, suggesting that interactions may play a significant role in triggering AGNs in LRDs.

In this paper, we take a step forward by performing a detailed analysis of {\it Virgil}, leveraging the very recent data collected by the MIDIS (PID: 6511, PI: {\"O}stlin), PAHSPECS (PID: 5279, PIs: Shivaei, D\'{i}az-Santos, and Boogaard), and OASIS (PID: 5997, PIs: Looser \& D'Eugenio) teams. These include exceptionally deep MIRI imaging at 5.6, 7.7, 10, and 15 $\mu$m, and deep NIRSpec/PRISM data. The deep NIRSpec/PRISM data enable the most comprehensive analysis of a LRD to date, providing critical insights into {\it Virgil}'s physical properties at UV and optical wavelengths, while the new MIRI data at 15 $\mu$m further confirm its extreme nature, suggesting a strong similarity to other EROs previously studied in the {\it Spitzer} era.

\vspace{3mm}
This paper is structured as follows. In Section 2, we describe the dataset used in this study, including HST and JWST observations, with particular focus on the MIRI data reduction. Section 3 presents a detailed analysis of the spectral properties of {\it Virgil}, based on NIRSpec data obtained by the OASIS team. Sections 4 and 5 present updated photometry and SED fitting, incorporating the newly acquired MIRI data from the MIDIS and PAHSPECS teams. In Section 6, we discuss the nature of {\it Virgil}, and in Section 7, we summarize our main conclusions.

\smallskip

Throughout this paper, we consider a cosmology with $H_{0} = 70\; \rm km\;s^{-1}\;Mpc^{-1}$, $\Omega_{M} = 0.3$, and $\Omega_{\Lambda} =0.7$. All magnitudes are total and refer to the AB system \citep{oke_secondary_1983}. A \citet{kroupa_variation_2001} initial mass function (IMF) is assumed (0.1--100 M$_{\odot}$).
Moreover, in Table \ref{tab:emission_ratios}, we report the line ratios that will be adopted throughout this work.

\begin{table}[h]
    \centering
    \renewcommand{\arraystretch}{1.5} 
    \begin{tabular}{l c}
        \hline
        Index & Definition \\
        \hline
        $\mathrm{R}3$ & $\log_{10} \left( \frac{ [\textsc{O iii}]\lambda5007 } {\textsc{H}\beta} \right)$ \\
        $\mathrm{O}32$ & $\log_{10} \left( \frac{ [\textsc{O iii}]\lambda\lambda 4959,5007 } { [\textsc{O ii}]\lambda\lambda 3727,3729} \right)$ \\
        $\mathrm{R}23$ & $\log_{10} \left( \frac{ [\textsc{O iii}]\lambda\lambda 4959,5007 + [\textsc{O ii}]\lambda\lambda 3727,3729}{\text{H}\beta} \right)$ \\
        $\mathrm{Ne}3\mathrm{O}2$ & $\log_{10} \left( \frac{ [\mathrm{Ne\,\textsc{iii}}]\lambda 3869 } { [\textsc{O ii}]\lambda\lambda 3727,3729} \right)$ \\
        $\mathrm{Ne}3\mathrm{O}2\mathrm{Hd}$ & $\log_{10} \left( \frac{ [\mathrm{Ne\,\textsc{iii}}]\lambda 3869 + [\textsc{O ii}]\lambda\lambda 3727,3729}{\text{H}\delta} \right)$ \\
        $\mathrm{O3Hg}$ & $\log_{10} \left( \frac{ [\textsc{O iii}]\lambda4363 } {\textsc{H}\gamma} \right)$ \\
        $\mathrm{O33}$ & $\log_{10} \left( \frac{ [\textsc{O iii}]\lambda5007 } { [\textsc{O iii}]\lambda4363 } \right)$ \\
        \hline
    \end{tabular}
    \caption{Definitions of line ratios adopted in this work. We use the O32 definition from \citet{calabro_evidence_2024}, that has a fixed offset of log$_{10}(1.3)$ between our values and the more common definition of O32 ($\equiv$ [O\,{\sc iii}]$\lambda$5007/[O\,{\sc ii}]$\lambda\lambda$ 3727,3729).}
    \label{tab:emission_ratios}
\end{table}

\section{Dataset}
In this work, we used data from HST and JWST, which are available in GOODS-S. In particular, we will briefly describe the data reduction for the MIDIS data and the NIRSpec/PRISM data collected by OASIS.

\subsection{HST}
Regarding the HST data, we used the ACS/WFC and WFC3/IR data from the Hubble Legacy Field (HLF) observations that cover GOODS-S. The HLF provides deep imaging in 9 HST bands covering a wide range of wavelengths (0.4$-$1.6 $\mu$m), from the optical (ACS/WFC F435W, F606W, F775W, F814W, and F850LP filters) to the near-infrared (WFC3/IR F105W, F125W, F140W and F160W filters). We refer the reader to \citet{whitaker_hubble_2019} for a more detailed description of this dataset.

\subsection{NIRCam}
We made use of NIRCam data from JADES Data Release 2 (JADES DR2 -- PIDs: 1180, 1210; PIs.: D. Eisenstein, N. Luetzgendorf; \citealt{eisenstein_jades_2023, eisenstein_overview_2023}), which includes observations from the JWST Extragalactic Medium-band Survey (JEMS -- PID: 1963; PIs: C. C. Williams, S. Tacchella, M. Maseda; \citealt{williams_jems_2023}) and the First Reionization Epoch Spectroscopically Complete Observations (FRESCO -- PID: 1895; PI: P. Oesch; \citealt{oesch_jwst_2023}).

The JADES/NIRCam data allowed us to cover a wide range in wavelengths ($\approx 1 - 5\mu\text{m}$) with the following bands: F090W, F115W, F150W, F182M, F200W, F210M, F277W, F335M, F356W, F430M, F444W, F460M, F480M. The estimated 5$\sigma$ depth from 30.5 to 30.9~mag (depths are calculated using 0.2\arcsec-diameter circular apertures assuming point-source morphologies; see \citealt{hainline_cosmos_2024} for details).

\subsection{MIRI}

We utilized MIRI data from the MIDIS (PID: 6511, PI: {\"O}stlin), PAHSPECS (PID: 5279, PIs: Shivaei, D\'{i}az-Santos, and Boogaard), and SMILES (PID: 1207, PI: Rieke) programs, which allowed us to cover a wide range of wavelengths ($5.6-25.5\;\mu$m). Specifically, MIDIS provides observations at 5.6, 7.7, and 10 $\mu$m, PAHSPECS enables photometry at 15 $\mu$m, and SMILES samples 12.8, 18, 21, and 25.5 $\mu$m. For SMILES, we relied on publicly available images released by the SMILES team. In contrast, we independently reduced the MIDIS data, while the PAHSPECS team provided a $5\arcsec \times 5\arcsec$ cutout centered on {\it Virgil}. Details on the MIRI/F1500W data reduction will be presented in Shivaei et al. (in prep.), based on the procedures described in \citet{perez-gonzalez_what_2024} and \citet{ostlin_miri_2024}.

To process the MIDIS data, we employed our custom MIRI pipeline (already calibrated in \citealt{rinaldi_midis_2023, iani_midis_2024-1} and further improved), built on the latest {\tt jwst} pipeline ({\tt 1.17.1}) with {\tt pmap\_1321}. We used the default {\tt stage 1} and {\tt stage 2} steps of the {\tt jwst} pipeline to produce calibrated data ({\tt *\_cal.fits}). Additionally, we implemented custom steps for creating a master background, performing a super-background subtraction and homogenization, and alignment to a common world coordinate system (WCS) based on the JADES DR2 catalog (\citealt{eisenstein_jades_2023}).

Our custom pipeline relies on a multi-tiered approach. First, it combines the {\tt *\_cal.fits} files into an initial mosaic using a custom drizzle algorithm based on the official package provided by STScI\footnote{\url{https://github.com/spacetelescope/drizzle}.}. This initial mosaic serves as the baseline for creating a source mask to exclude sources in the individual {\tt *\_cal.fits} files. To construct this mask, our pipeline uses {\sc Source Extractor} \citep{bertin_sextractor_1996}. A preliminary {\tt master\_background} is then built by reprojecting this preliminary source mask onto each exposure, stacking all exposures into a 3D array, and computing the {\tt biweight location} along the z-axis. This method minimizes the influence of outliers and provides a robust estimate of the 2D background (i.e., {\tt master\_background}).

For each exposure, our pipeline iteratively estimates a scale factor ({\tt A}) to optimize the subtraction of the {\tt master\_background}. The scale factor is recalculated at each iteration using clipped data to ensure convergence and minimize residuals between the corrected and original images. Once the {\tt master\_background} is subtracted, the pipeline filters each exposure along rows and columns to remove residual patterns while masking out the sources. This procedure is similar to the standard approach used to mitigate wisp effects in NIRCam data, as described by \citet{bagley_next_2024}.

Then, the pipeline drizzles together all the single exposures to produce a new mosaic, from which it creates a {\tt master\_source\_mask}. At this level, the pipeline makes use of both {\sc SourceExtractor} and {\sc NoiseChisel} \citep{akhlaghi_noise-based_2017} to create the {\tt master\_source\_mask}: {\sc SourceExtractor} effectively identifies compact sources. In contrast {\sc NoiseChisel} detects faint sources and extended regions around bright objects. The {\tt master\_source\_mask} is then reprojected onto each single exposure to improve the {\tt master\_background} and super-background subtraction and homogenization, following a methodology similar to the one outlined in \citet{bagley_next_2024}.

After this step, the exposures are filtered again in rows and columns and aligned to the JADES DR2 reference catalog using a modified version of the {\tt tweakreg\_step} within the {\tt jwst} pipeline.

The final steps include running the {\tt outlier\_step} and {\tt resampling\_step}, producing the final mosaic. This final mosaic is then used to refine the {\tt master\_source\_mask} and iterate the entire procedure for further optimization. Similar methodologies have already been widely used when reducing MIRI data (e.g.,  \citealt{alberts_smiles_2024, perez-gonzalez_what_2024, ostlin_miri_2024}).

Finally, we estimated the 5$\sigma$ depth in our reduced MIRI images, which is 28.6, 28.3, 27.1, and 24.9 mag at, respectively, 5.6, 7.7, 10, and 15 $\mu$m ($r = 0.3$\arcsec). The 5$\sigma$ depth for the SMILES images at 12.8, 18, 21, and 25.5 $\mu$m is 24.4, 23.3, 22.8, and 20.8 mag (\citealt{alberts_smiles_2024}; $r = 0.4 - 0.7$\arcsec). Depths are estimated using circular apertures assuming point-source morphologies.

\subsection{NIRSpec/PRISM}

The OASIS program consists of two ultra-deep NIRSpec/MSA pointings, each with a total exposure time of 28 hours, using the PRISM/CLEAR configuration to obtain low-resolution ($R\approx30\text{--}300$) spectroscopy over the $0.6 - 5.3\,\mu$m range. The observations are executed in multiple visits, with each pointing incorporating three nodding positions within the MSA to improve background subtraction, and multiple dithered observations to mitigate detector artifacts. The MSA configuration is designed to target a sample of 223 galaxies at $5 < z < 8$. The NIRSpec observations are performed using the NRSRAID6 readout pattern with three groups per integration and a total of six integrations per exposure.

To reduce the NIRSpec/PRISM data, we made use of the pipeline developed by the ESA NIRSpec Science Operations Team (SOT) and the Guaranteed Time Observations (GTO) NIRSpec teams, with optimizations aimed at improving background subtraction, rectification, 1D extraction, and spectral combination. While broadly consistent with the STScI pipeline used for generating archive products, these enhancements ensure more accurate spectral calibration, particularly for compact sources affected by spatial offsets within the slit. A wavelength correction is applied to mitigate this bias, which arises when sources are not perfectly centered along the dispersion axis. For a more detailed description of the NIRSpec data reduction, we refer the reader to \citet{bunker_jades_2024} and \citet{deugenio_jades_2024}. A more detailed description of this dataset will be presented in Looser \& D'Eugenio et al. (in prep.).

\section{On the nature of Virgil: Insights from NIRSpec/PRISM}
In this section, we delve into the nature of {\it Virgil} based on  NIRSpec/PRISM data. Due to the extended nature of Virgil in the NIRSpec wavelength range, we corrected the flux following the method outlined in \citet{carniani_spectroscopic_2024}. We applied a first-order polynomial ($\alpha_{1} + \alpha_{0}\lambda$) to model the wavelength-dependent slit losses and calibrated the flux to match the NIRCam photometry. That is, we focus on the rest-frame UV and optical spectral regions. All the line fluxes and properties estimated in this section are summarized in Table \ref{tab:galaxy_em_prop}.

\begin{deluxetable}{lcc}
\tablecaption{Emission Line Fluxes and Derived Properties for {\it Virgil}}
\tablewidth{0pt}
\tablehead{
\colhead{Line} & \colhead{Value} & \colhead{Error}
}
\startdata
N\,{\sc iv}]\,$\lambda$1487 & 47.3 & 19.0 \\
C\,{\sc iv}\,$\lambda$1549 & 28.3 & 18.7 \\
C\,{\sc iii}]\,$\lambda$1906 & 48.4 & 16.4 \\
{[O\,{\sc ii}]}\,$\lambda\lambda$3727,3729 & 67.4 & 5.4 \\
{[Ne\,{\sc iii}]}\,$\lambda$3867 & 34.6 & 5.5 \\
He\,{\sc i}\,$\lambda$3889 & 20.9 & 5.5 \\
{[Ne\,{\sc iii}]}\,$\lambda$3968 & 22.9 & 4.5 \\
H$\delta$\,$\lambda$4102 & 20.5 & 4.2 \\
H$\gamma$\,$\lambda$4340 & 32.4 & 4.9 \\
{[O\,{\sc iii}]}\,$\lambda$4363 & 16.5 & 4.7 \\
H$\beta$\,$\lambda$4861 & 92.5 & 4.1 \\
{[O\,{\sc iii}]}\,$\lambda$4959 & 206.0 & 5.2 \\
{[O\,{\sc iii}]}\,$\lambda$5007 & 587.9 & 5.2 \\
He\,{\sc i}\,$\lambda$5877 & 15.5 & 4.2 \\
H$\alpha$\,$\lambda$6563$^{*}$ & 304.9 & 6.6 \\
\hline
H$\alpha$~(narrow)\,$\lambda$6563 & 219.9 & 12.5 \\
H$\alpha$~(broad)\,$\lambda$6563 & 168.8 & 17.9 \\
\hline
\hline
Galaxy Properties & & \\
\hline
\hline
FWHM (broad H$\alpha$) [km/s] & 1750 & 575 \\
$E(B-V)_{gas}$ [mag] & 0.24 & 0.07 \\
$E(B-V)_{stars}$ [mag] & 0.10 & 0.03 \\
$A_V$ (H$\alpha$/H$\beta$) [mag] & 0.65 & 0.18 \\
12+log(O/H)$^{**}$ & 7.78 & 0.18 \\
$T_{e}(O^{+})$ [K] & 14800 & 600 \\
$T_{e}(O^{2+})$ [K] & 18300 & 3200 \\
$\text{log}_{10}(\mathcal{U})^{**}$ & -2.03 & 0.08 \\
$n_{e}$$^{***}$ [cm$^{-3}$] & $<500$ & -- \\
$\text{EW}_{0}(\text{H}\alpha)$ [\AA] & 422 & 20 \\
$\text{EW}_{0}$([O\,{\sc iii}]$\lambda\lambda4959.5007$) [\AA] & 1514 & 18 \\
SFR(H$\alpha$) [$M_\odot\,\mathrm{yr}^{-1}$] & 6.10 & 0.93 \\
SFR(UV) [$M_\odot\,\mathrm{yr}^{-1}$] & 2.96 & 1.02 \\
$\Sigma_{SFR(\text{H}\alpha)}$ [$M_\odot\,\mathrm{yr}^{-1}\,\mathrm{kpc}^{-2}$] & 5.25 & 0.80 \\
$\mathcal{B}$ & 0.32 & 0.16 \\
$M_{UV}$ [mag] & -18.70 & 0.17 \\
$\beta$ & -1.76 & 0.18 \\
$f_{esc,LyC}$ & 0.02 & 0.01 \\
$\xi_{ion}$ \, $[\log_{10}(\mathrm{Hz\, erg}^{-1})]$ & 24.91 & 0.18 \\
\hline
$z_{spec}$ & 6.6379 & 0.0035 \\
\enddata
\tablecomments{Observed fluxes and errors (not corrected for dust) are given in units of $10^{-20} \, \mathrm{erg\,s^{-1}\,cm^{-2}}$. We also show the H$\alpha$ fluxes for the narrow and broad components obtained by {\sc MSAEXP}. Derived galaxy properties with associated uncertainties are provided. $^{*}$ refers to H$\alpha$ corrected for the contribution from [N\,{\sc ii}]$\lambda\lambda6548,6583$. $^{**}$ refers to the estimated median quantities, while $^{***}$ refers to upper limits ($2\sigma$).}
\label{tab:galaxy_em_prop}
\end{deluxetable}

\subsection{Redshift assessment of Virgil}


\begin{figure*}
    \centering
    \includegraphics[width=0.99\linewidth]{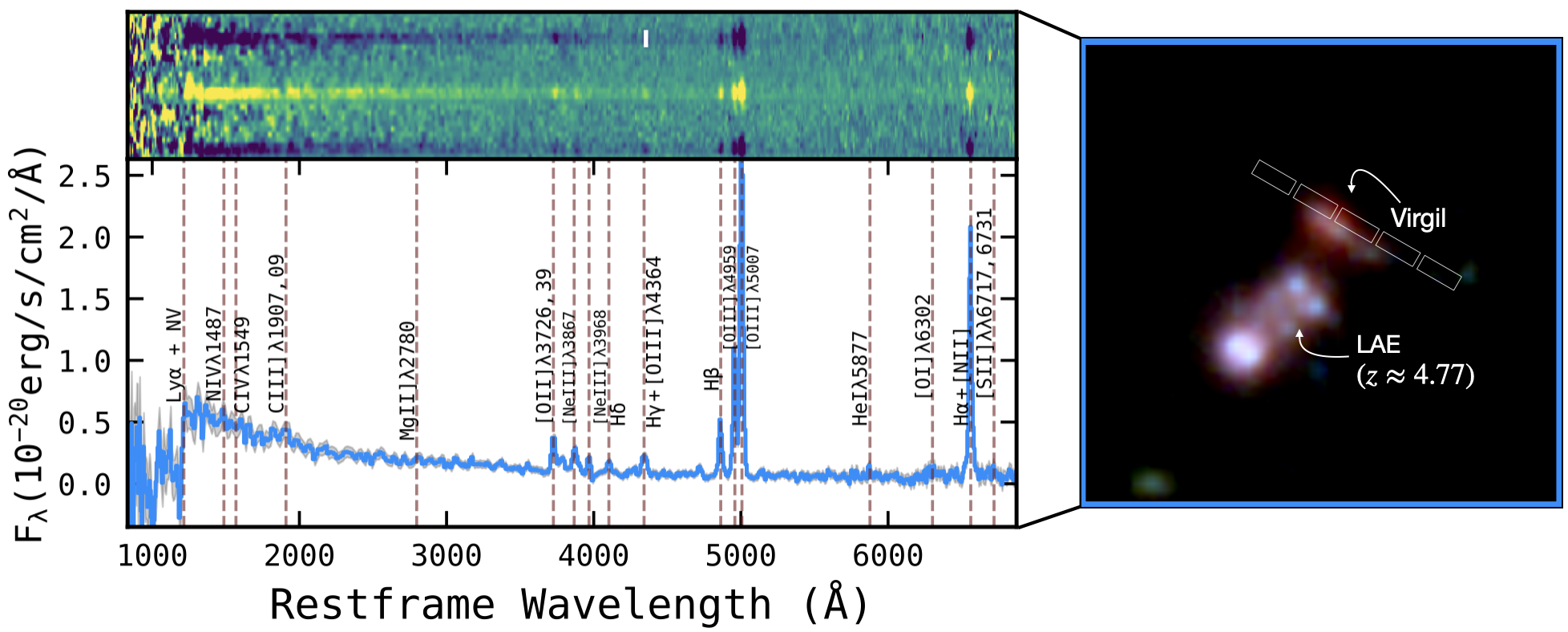}
    \caption{
    {\bf Left panel:} 2D MSA/PRISM spectrum produced by {\sc MSAEXP}. We optimally scaled the trace to highlight all of the significant line detections. We show the data in azure, while the uncertainty is highlighted in gray. Assuming the best-fit {\sc MSAEXP} $z_{spec} = 6.6379\pm0.0035$, we show the positions of the emission lines as dashed vertical lines. {\bf Right panel:} JWST/NIRCam RGB image (2.5\arcsec$\times$2.5\arcsec) along with the slit positions showed in white. We highlight {\it Virgil} and the foreground LAE at $z\approx4.77$ already studied in \citet{matthee_deciphering_2022} with VLT/MUSE data.}
    \label{fig:virgil_spectrum}
\end{figure*}

We used {\sc MSAEXP} \citep{brammer_msaexp_2023, heintz_jwst-primal_2025} to estimate the spectroscopic redshift. {\sc MSAEXP} fits the data using a combination of Gaussian continuum splines and emission line templates. We set {\tt nsplines = 11} as an input parameter and searched for the minimum reduced $\chi^{2}$ value within the redshift range $z = 6-7$, based on the prior spectroscopic estimate from VLT/MUSE (\citealt{bacon_muse_2023}).
As shown in the 2D spectrum (Figure~\ref{fig:virgil_spectrum}), a faint Ly$\alpha$ feature—possibly blended with NV$\lambda1240$—is visible, consistent with the detection reported by \citet{bacon_muse_2023}. The line appears spatially offset, suggesting it may originate from a distinct kinematic component, potentially explaining the slight redshift discrepancy with the MUSE measurement. Several other emission lines are detected, yielding an unambiguous redshift of $z_{\mathrm{spec}} = 6.6379 \pm 0.0035$, which is slightly higher than the VLT/MUSE estimate\footnote{The redshift from VLT/MUSE is based on a single line identification.} but still consistent to the second decimal place.

\subsection{Virgil's physical properties from its UV/optical spectrum}

\subsubsection{Balmer Decrement and Dust Content}
We estimated the color excess, $E(B-V)_{gas}$, by analyzing the Balmer decrement between H$\beta$ and H$\alpha$. We measured an observed Balmer decrement of $3.70 \pm 0.19$. However, since we cannot resolve [N\,{\sc ii}]$\lambda\lambda$6548, 6583 individually, we applied the correction from \citet{anders_spectral_2003} for low-metallicity objects ($10-20\%\,Z_{\odot}$; see next section), where the corrected H$\alpha$ flux is $\approx91\%$ of the observed (H$\alpha$ + [N\,{\sc ii}]$\lambda\lambda$6548, 6583). After this correction, the Balmer decrement is $3.30 \pm 0.16$. Assuming Case B recombination \citep{osterbrock_astrophysics_2006}, and adopting an electron temperature of $1.8 \times 10^4$~K (as derived in the next section) and an electron density of $10^2$~cm$^{-3}$, we use \textsc{PyNeb} \citep{luridiana_pyneb_2015} to compute an intrinsic H$\alpha$/H$\beta$ ratio of 2.76. The measured ratio corresponds to $E(B-V)_{gas} = 0.24 \pm 0.07$~mag, assuming the Small Magellanic Cloud (SMC) reddening law \citep{gordon_quantitative_2003}, which has been shown to reproduce well the dust attenuation in high-redshift galaxies \citep[e.g.,][]{reddy_mosdef_2018} and reddened quasars \citep[e.g.,][]{hopkins_dust_2004}. This value leads to an attenuation of $A_V = 0.65 \pm 0.18$~mag, adopting $R_V = 2.74 \pm 0.13$ \citep{gordon_quantitative_2003}. It further implies $E(B-V)_{stars} = 0.10 \pm 0.03$~mag\footnote{Assuming the differential dust attenuation, we adopt $E(B-V)_{stars} = 0.44 \times E(B-V)_{gas}$ \citep{koyama_different_2019, shivaei_mosdef_2020}.}.

To further assess the dust attenuation, we also compared the Balmer decrement for H$\beta$ with respect to H$\gamma$ and H$\delta$ (expected theoretical ratios of 0.474 and 0.263, respectively, under Case B recombination; \citealt{osterbrock_astrophysics_2006}). The resulting $E(B-V)_{\text{gas}}$ values were consistent with a low level or zero attenuation.

If \textit{Virgil} instead hosted a Type 1 AGN (see Section 3.4), the broad component of H$\alpha$ must be excluded when computing the Balmer decrement. In this case, considering only the narrow component of H$\alpha$ and treating the detected broad H$\beta$ as a narrow line due to its low significance, we would obtain a lower Balmer decrement of $2.38 \pm 0.18$, consistent with a dust-free scenario. Alternatively, in this scenario, the Balmer decrement may be intrinsically lower, as observed in some $z \approx 4-7$ galaxies with density-bounded regions or in low-mass extreme emitters \citep{sandles_jades_2024, mcclymont_density-bounded_2024, scarlata_universal_2024}.

\subsubsection{Gas-Phase Metallicity and Ionization State}

\begin{figure}
    \centering
    \includegraphics[width=1.\linewidth]{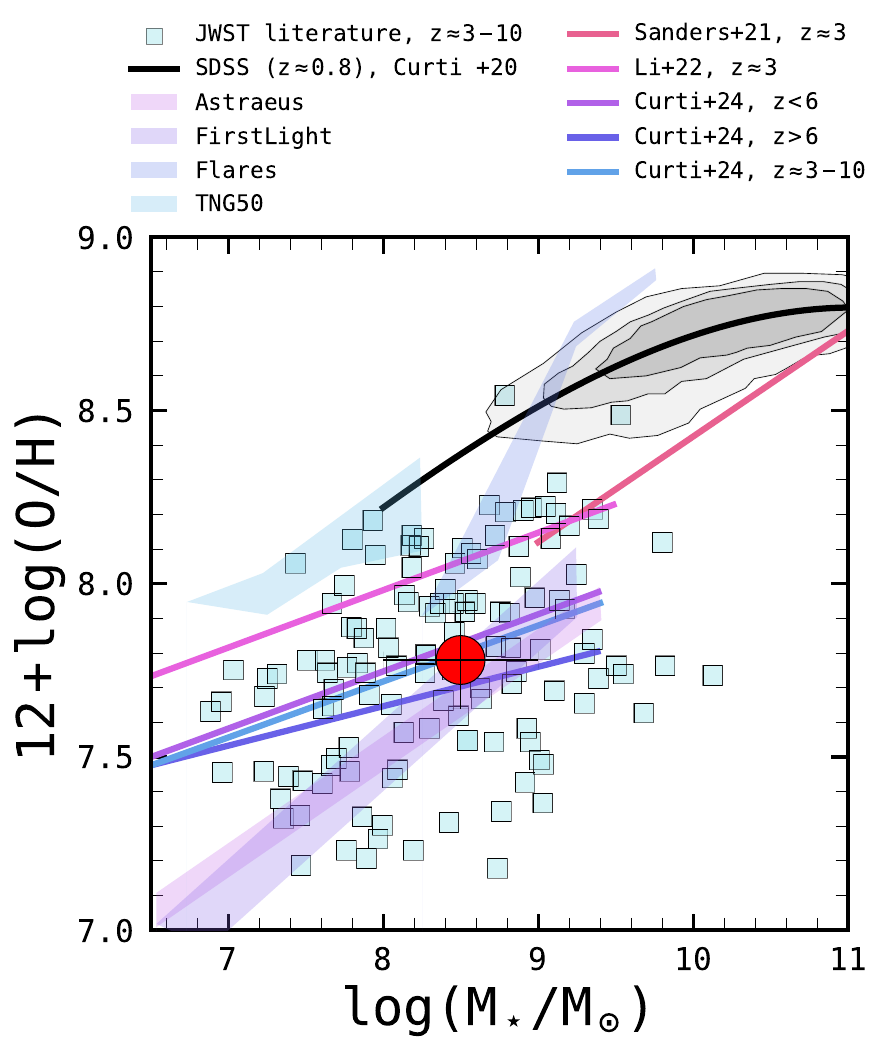}
    \caption{The stellar mass-gas-phase metallicity relation. {\it Virgil} is shown as a red filled circle. The stellar mass is taken from \citet{iani_midis_2024} (median value), with an error bar reflecting the uncertainty in the estimate, consistent with the results obtained in this paper (see Section~5). Filled squares represent recent JWST measurements at $z \approx 3$--$10$, including galaxies from the JADES field \citep{curti_jades_2024}, the CEERS field \citep{nakajima_jwst_2023}, and the SMACS J0723.3-7327 sample \citep{curti_chemical_2023}. For comparison, we show MZR determinations at lower redshifts from \citet{curti_mass-metallicity_2020} (Sloan Digital Sky Survey; SDSS) and \citet{sanders_mosdef_2021} (MOSDEF at $z \approx 2$--$3$). Additionally, we include the best-fit of the low-mass end of the MZR at $z \approx 3$ from \citet{li_mass-metallicity_2023}, based on JWST/NIRISS slitless spectroscopy. Finally, we also show theoretical predictions at high redshift from {\sc Astraeus} (\citealt{ucci_astraeus_2023}), {\sc FirstLight}  (\citealt{langan_weak_2020}), the First Light and Reionisation Epoch Simulations ({\sc FLARES}; \citealt{lovell_first_2021}), and {\sc TNG50} simulations (\citealt{nelson_first_2019}).}
    \label{fig:gas_phase_metallicity}
\end{figure}

Both \textsc{MSAEXP} and our internal routine successfully deblended H$\gamma$ and [O\,\textsc{iii}]$\lambda4363$. Additionally, we detect the [O\,\textsc{ii}]$\lambda\lambda3727,3729$ doublet. These lines, along with [O\,\textsc{iii}]$\lambda\lambda4959,5007$,  enabled us to estimate the electron temperatures ($T_e$) of O$^+$ and O$^{2+}$ and the gas-phase metallicity (e.g., \citealt{izotov_chemical_2006, curti_jades_2024, sanders_direct_2024}). For this purpose, we employed a non-parametric approach based on \citet{langeroodi_genesis-metallicity_2024}, utilizing the {\sc genesis-metallicity} tool\footnote{The tool is available at \url{https://github.com/langeroodi/genesis_metallicity/tree/main}.}, which is built on {\sc PyNEB} \citep{luridiana_pyneb_2015}. We derived $T_e$(O$^+$) $= (1.48 \pm 0.06) \times 10^4$~K and $T_e$(O$^{2+}$) $= (1.83 \pm 0.32) \times 10^4$~K (adopting the direct method). We highlight that $T_e$(O$^{2+}$) is higher than typical values observed in galactic H\;{\sc ii} regions, but is consistent with the temperatures measured in the most metal-poor galaxies (e.g., \citealt{skillman_alfalfa_2013}).

Furthermore, the {\sc genesis-metallicity} tool offers the option to estimate gas-phase metallicity using either the direct ($T_e$) method or strong-line calibrations, depending on the available line fluxes. Both approaches yielded consistently low metallicities for \textit{Virgil}, with $12 + \log(\mathrm{O/H}) = 7.70 \pm 0.14$ for the direct method, and $12 + \log(\mathrm{O/H}) = 7.91 \pm 0.04$ for the strong-line method. Based on the commonly accepted solar oxygen abundance (12 + log(O/H) = 8.69; \citealt{asplund_chemical_2009}), this gives us $Z/Z_{\odot} \approx 0.11-0.18$.   

In the recent literature, other emission lines have been proposed to trace the gas-phase metallicity (mostly from UV), such as C\,\textsc{iv}] $\lambda$1548 and C\,\textsc{iii}]$\lambda$1907 (see \citealt{mingozzi_classy_2022}). Based on their Equation 7, which exploits the equivalent width (EW) of C\,\textsc{iii}]$\lambda1907,09$, we estimated that $12 + \log(\mathrm{O/H}) = 7.78\pm0.21$ (in agreement with our previous measurements). As shown in the mass-metallicity relation (MZR; Figure \ref{fig:gas_phase_metallicity}), {\it Virgil} is in good agreement with the relation recently reported by \citet{curti_jades_2024} for galaxies at $z\gtrsim6$ by leveraging the JADES sample as well as the recent literature from JWST.


\smallskip

We further investigate the ionization state of {\it Virgil}'s gas by examining the ionization parameter, log$_{10}(\mathcal{U})$, a key ISM property that reflects the intensity and hardness of the ionizing radiation, gas density, and the geometry of gas relative to ionizing sources. The O32 index, a well-established tracer of log$_{10}(\mathcal{U})$, yields a value of $-2.03 \pm 0.08$ based on the empirical calibration from \citet{papovich_clear_2022} (their Equation 4), which was derived from galaxies at $1.1<z<2.3$ in the CLEAR survey \citep{simons_clear_2023}. Notably, this value is consistent with the estimate from UV-based tracers identified by \citet{mingozzi_classy_2022}, who calibrated log(C\,\textsc{iii}]/O\,\textsc{iii}]), log(C\,\textsc{iv}/C\,\textsc{iii}]), log(EW(O\,\textsc{iii}])), and log(EW(C\,\textsc{iv})) against O32. Applying their Equations 8 and 10, we find an average log$_{10}(\mathcal{U}) = -1.99 \pm 0.15$, confirming a high level of ionization in agreement with the O32-based result. Alternatively, using the relation between log$_{10}(\mathcal{U})$ and the Ne3O2 index from \citet{witstok_assessing_2021} (their Equation 3), we derive log$_{10}(\mathcal{U}) = -2.33 \pm 0.07$. While slightly lower than the previous estimates, this value remains indicative of a highly ionized ISM, which is in line with the average galaxy population during EoR.

We were also able to fit the [S\,\textsc{ii}]$\lambda\lambda6717,6731$ doublet, although with a low SNR ($\approx 2\sigma$). Following the theoretical framework of \citet{kewley_theoretical_2019}, we estimated an upper limit on the electron density, $n_e \approx 500\;\text{cm}^{-3}$, considering its weak dependence on $T_e$ \citep{zhang_are_2024}. However, the low spectral resolution of NIRSpec/PRISM\footnote{$R \approx 30$--$420$, depending on the observed wavelength.} prevents deblending the [O\,\textsc{ii}]$\lambda\lambda3727,3729$ doublet, limiting further constraints on $n_e$. Recently, \citet{topping_aurora_2025} analyzed deep JWST/NIRSpec spectra of SFGs at $z \approx 1.4 - 10$ as part of the AURORA survey \citep{shapley_aurora_2024}, focusing on the [S\,\textsc{ii}]$\lambda\lambda6717,6731$ doublet (detected up to $z\approx6.8$), its evolution over cosmic time, and its correlation with various parameters (e.g., line ratios, star formation rate). Notably, our upper limit aligns with the average $n_e$ expected at $z \approx 6.5$, and it is consistent with the estimate obtained using their relation with the star formation rate (SFR; their Equation 2) and Ne3O2 (their Equation 3).

\medskip
As a final note, we report the detection of the N\,\textsc{iv}]$\lambda1487$ emission line, while N\,\textsc{iii}]$\lambda1750$ is detected at a significance well below $2\sigma$. Additionally, {\sc MSAEXP} identifies a possible N\,\textsc{v}$\lambda1240$ feature, although it may be blended with Ly$\alpha$ due to the low resolution of NIRSpec/PRISM. The presence of N\,\textsc{iv}]$\lambda1487$ has been proposed as a signature of low-luminosity quasars \citep{glikman_discovery_2007, bunker_jades_2023, cameron_jades_2023}. However, we note that NIRSpec spectra often exhibit spurious oscillations in this wavelength range, and higher-resolution, deeper observations will be necessary to confirm the reality of these lines. For this reason, we do not discuss this further.

\subsubsection{UV-$\beta$ slope, Burstiness, and Ionizing Photon Production Efficiency}

We modified the official {\sc MSAEXP} code to enable estimation of the UV $\beta$ slope (in the rest-frame wavelength range $\lambda \approx 1500-2500$~\AA) by fitting a linear relation ($\log_{10}(f_{\lambda}) = \beta\,\text{log}_{10}\lambda + q$)
using non-linear least squares optimization via {\tt curve\_fit} ({\sc SciPy}). We also employed a Monte Carlo approach, performing 1000 linear fits to spectral realizations with randomly perturbed fluxes and adopting the median and standard deviation of the resulting $\beta$ values. We retrieved $\beta = -1.76 \pm 0.18$.


To investigate the potential role of \textit{Virgil} during EoR, we estimated its ionizing photon production efficiency, $\xi_{\text{ion}}$. Following the method outlined in \citet{rinaldi_midis_2024}, we first derived the Lyman continuum escape fraction ($f_{\text{esc, LyC}}$) from the observed UV $\beta$ slope using the prescription of \citet{chisholm_far-ultraviolet_2022}. We then computed $\xi_{\text{ion}}$ based on its dependence on $(1 - f_{\text{esc, LyC}})^{-1}$ (Equation 2 in \citealt{rinaldi_midis_2024}). We find $f_{esc, LyC} = 0.02 \pm 0.01$\footnote{Although the systematic uncertainty reported by \citet{chisholm_far-ultraviolet_2022} is on the order of 0.05, the lower bound of our estimate makes $f_{esc, LyC}$ consistent with 0\%. Nonetheless, the resulting $\xi_{ion}$ would not differ significantly from the reported value.} and $\log_{10}(\xi_{ion}/\text{Hz erg}^{-1}) = 24.91 \pm 0.18$. The latter is below the canonical value ($25.2\pm0.1$; \citealt{robertson_new_2013}), but still in agreement with the average high-$z$ population at $z\approx6-7$ (\citealt{matthee_production_2017, rinaldi_midis_2024, simmonds_ionising_2024}). This result is not a surprise, as the rest-frame equivalent width (EW$_{0}$) for H$\alpha$ is about $422\pm20$ \AA, and very high $\xi_{ion}$ is often associated with very strong line emitters (i.e., large EW$_{0}$(H$\alpha$); see Figure 5 from \citealt{rinaldi_midis_2024}).

Following the work presented in \citet{atek_star_2022} and \citet{navarro-carrera_burstiness_2024}, high-$z$ galaxies are expected to grow through bursty episodes of star formation, driven by large temporal fluctuations in their SFR—a trend recently suggested also by \citet{langeroodi_nirspec_2024}.

The NIRSpec/PRISM spectrum enabled us to estimate the star formation rate (SFR)\footnote{Both SFR(H$\alpha$) and SFR(UV) are corrected for dust attenuation.} directly from emission lines, specifically H$\alpha$, using the calibration from \citet{kennicutt_star_2012} based on a Kroupa IMF. In particular, we adopt:

\begin{equation}
    \log_{10} \left(\frac{\mathrm{SFR}}{M_{\odot}\,\mathrm{yr}^{-1}} \right) = \log_{10} \left(L_{\mathrm{H\alpha}}/(\text{erg}\,\text{s}^{-1})\right) - 41.64,
\end{equation}

\noindent
where the coefficient 41.64 corresponds to the calibration at sub-solar metallicity (10$-$20\% $Z_\odot$; \citealt{theios_dust_2019}). 

We find that log$_{10}$(SFR(H$\alpha$)/($M_{\odot}\;yr^{-1}$)) $=0.79\pm0.07$. Based on the estimate from \citet{iani_midis_2024} for the effective radius ($r_{\text{eff}} \approx 0.43$~pkpc), we retrieve $\Sigma_{SFR(\text{H}\alpha)}= \text{SFR}(\text{H}\alpha)/2\pi r_{\text{eff}}^{2} = 5.25\pm0.80\;M_{\odot}\,yr^{-1}\,\text{kpc}^{-2}$, which is in line with recent results from \citet{calabro_evolution_2024} at $z\approx6.5$.

The detection of H$\alpha$ allowed us to estimate another key parameter: the burstiness, defined as $\mathcal{B} = \log_{10}(\mathrm{SFR}(\mathrm{H}\alpha)/\mathrm{SFR}(\mathrm{UV}))$ \citep{atek_star_2022}. We derived $\mathrm{SFR}(\mathrm{UV})$ at 1500~\AA\ using the calibration from \citet{kennicutt_star_2012}, adapted for sub-solar metallicity following \citet{theios_dust_2019}, obtaining $\log_{10}\left(\mathrm{SFR}(\mathrm{UV})/M_{\odot}\,\mathrm{yr}^{-1}\right) = 0.47 \pm 0.15$. This yields a burstiness parameter of $\mathcal{B} = 0.32 \pm 0.16$, indicating that the source is currently forming stars at a rate more than two times higher than its recent past average.

The timescales of H$\alpha$ and UV emission provide further insights. Following the discussion presented in \citet{faisst_recent_2019}, \citet{iani_midis_2024-1}, and \citet{navarro-carrera_burstiness_2024}\footnote{See Figure 12 in \citealt{faisst_recent_2019} based on different SFHs, Figure 15 in \citet{iani_midis_2024-1} based on {\sc starburst99}, and Figure 7 in \citealt{navarro-carrera_burstiness_2024} based on {\sc FIRSTLIGHT} simulations.}, after a star formation episode, H$\alpha$ is initially enhanced due to newly formed O and B stars, while the UV luminosity is still rising. This results in SFR(H$\alpha$)/SFR(UV) $>1$, which has been suggested as a signature of a bursty star formation phase \citep{atek_star_2022}. As massive stars rapidly evolve and die, H$\alpha$ emission declines, while longer-lived UV-emitting stars sustain the UV luminosity, leading to SFR(H$\alpha$)/SFR(UV) $<1$. As these UV-bright stars also fade, the ratio eventually returns to unity.

Our findings suggest that {\it Virgil} likely experienced a previous burst of star formation well before the most recent one (i.e., the one currently producing H$\alpha$), which explains its SFR(UV) and likely placed its $\mathcal{B}$ below unity. We now measure $\mathcal{B} =0.32 \pm 0.16$, indicating that the source is either entering a new bursty phase or fading out of one. Notably, this is consistent with its position in the $\mathrm{SFR}$–$M_{\star}$ plane \citep{rinaldi_galaxy_2022, rinaldi_emergence_2024}, where it overlaps with the separation line for the starburst cloud proposed by \citet{caputi_star_2017, caputi_alma_2021}. This suggests we may be observing {\it Virgil} at a transitional stage in its star formation history.

\subsection{The ISM Conditions of Virgil: Excitation vs. Ionization}

\begin{figure*}
    \centering
    \includegraphics[width=0.49\linewidth]{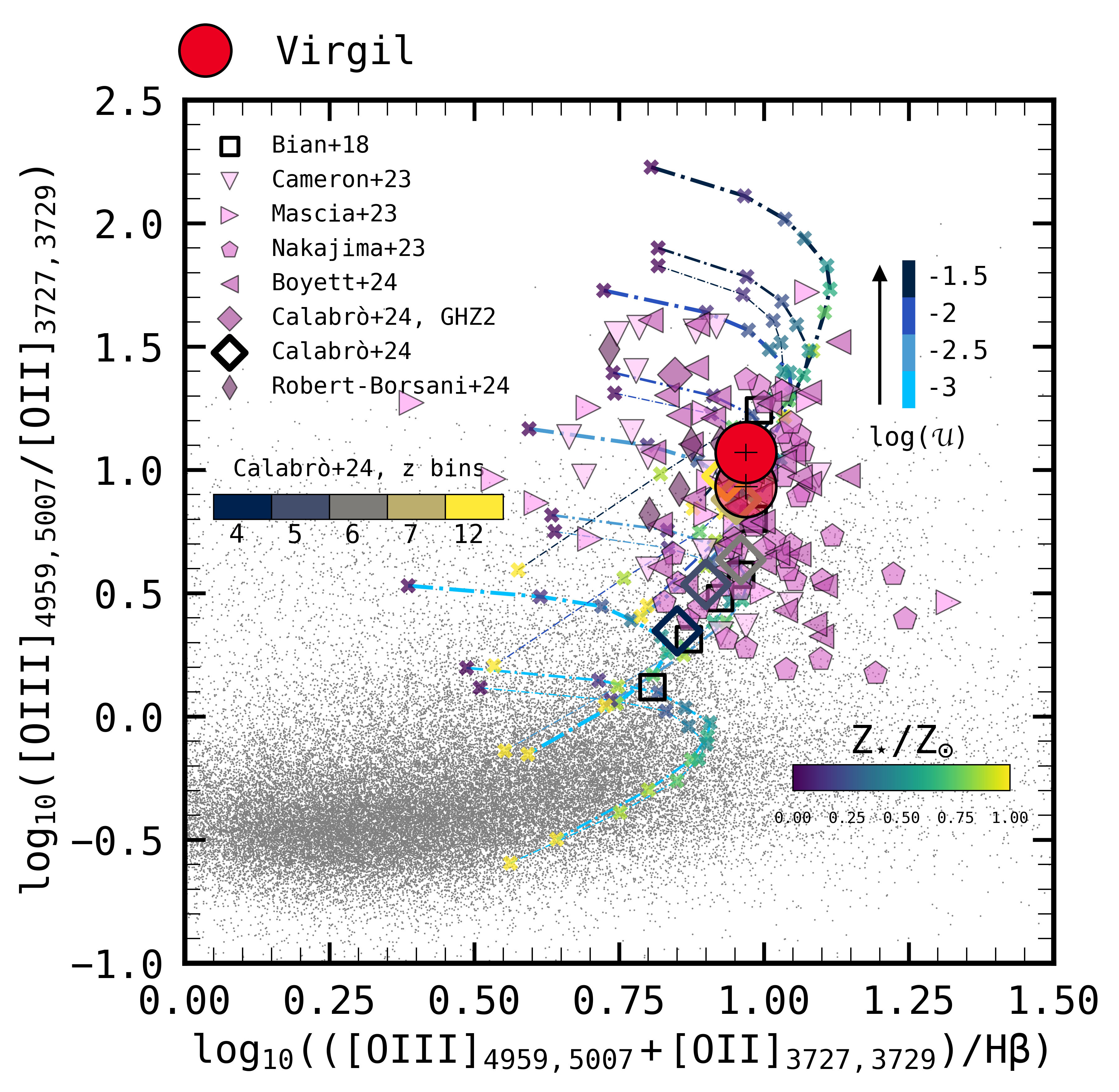}
    \includegraphics[width=0.49\linewidth]{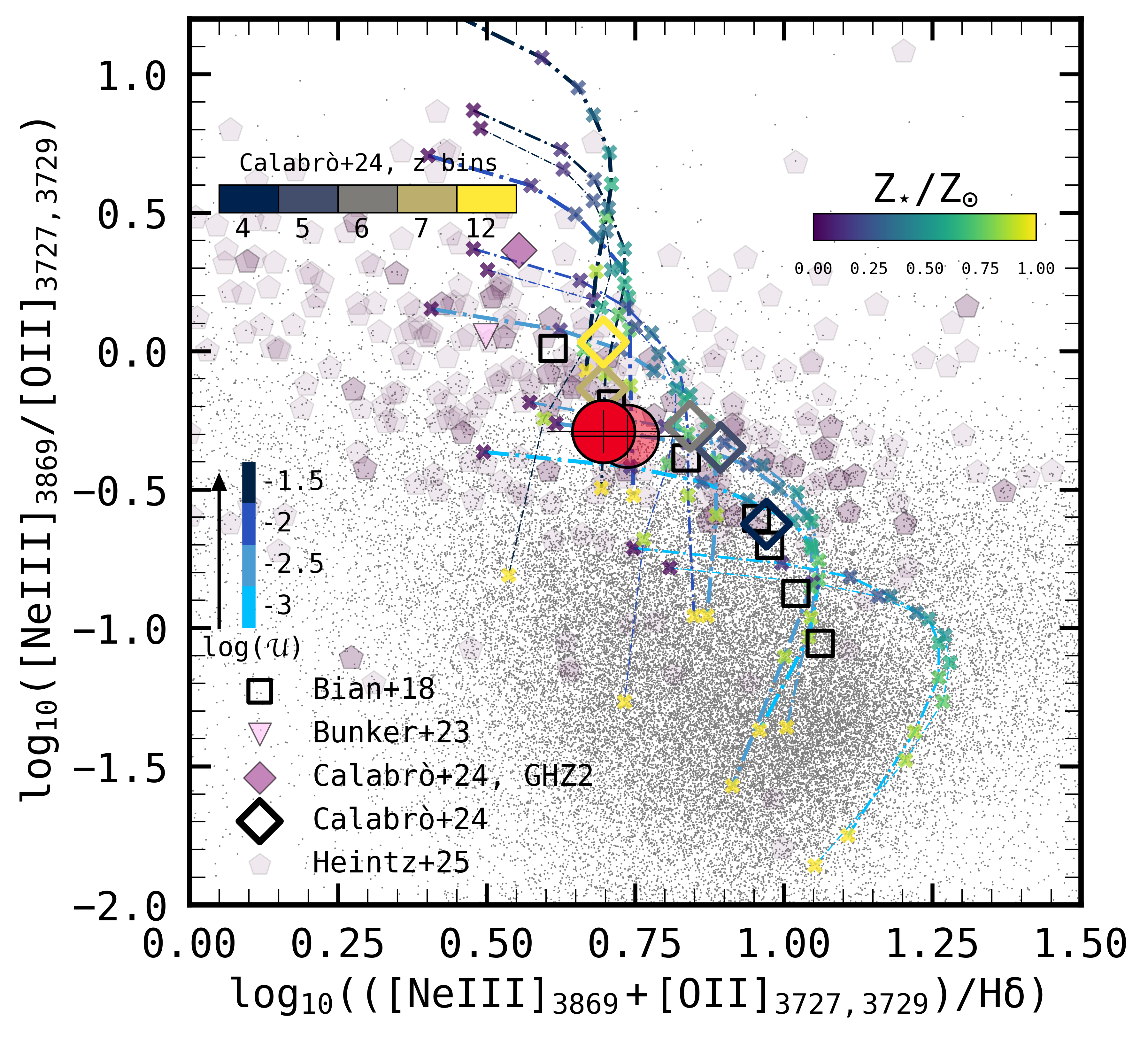}
    \caption{{\bf Left Panel:} O32$-$R23. {\bf Right Panel:} Ne3O2Hd$-$Ne3O2. {\it Virgil} is shown as a red filled circle, with higher opacity representing the case without dust correction and lower opacity indicating the dust-corrected one. In both panels, we show the photionization models from {\sc CLOUDY} with the following setup: $n_{e}$ (10$^{2}$, 10$^{3}$, and 10$^{4}$ cm$^{-3}$; increasing thickness), log$_{10}(\mathcal{U})$ (-3, -2.5, -2, -1.5; different shades of blue), and different metallicities (0.05 to solar; color-coded in viridis using “x” symbols). For comparison, we show low-$z$ SF galaxies from SDSS (\citealt{york_sloan_2000, kauffmann_stellar_2003}) and local high-$z$ analogues from \citet{bian_direct_2018}. To put our results in context, we also show recent observations at high-$z$ from \citet{bunker_jades_2023, cameron_jades_2023, mascia_closing_2023, nakajima_jwst_2023, boyett_extreme_2024, calabro_evidence_2024, calabro_evolution_2024, roberts-borsani_between_2024, heintz_jwst-primal_2025} at $z\approx4-12$.}
    \label{fig:ism}
\end{figure*}

Since the NIRSpec/PRISM spectrum provides access to [O\,{\sc ii}]$\lambda\lambda$3727,3729, [Ne\,{\sc iii}]$\lambda$3869, H$\delta$, H$\beta$, and [O\,{\sc iii}]$\lambda\lambda$4959,5007 at $z\approx6.64$, we investigated the interstellar medium (ISM) conditions of {\it Virgil} by leveraging the following line ratios: O32, R23, Ne3O2, and Ne3O2Hd. 
To fully explore the complexity of the ISM in this source, we closely followed the methodology described in \citet{calabro_near-infrared_2023} and \citet{calabro_evidence_2024}, running {\sc pyCloudy} (v0.9.15) with CLOUDY version 17.01 \citep{ferland_2017_2017}. We adopted their setup (see \citealt{calabro_evidence_2024}, Section 2.3) to generate line predictions for direct comparison with our results.

Briefly, we modeled SFGs with a spherically symmetric, radiation-bounded shell of gas surrounding a population of young (O- and B-type) stars, adopting the incident radiation field from BPASS stellar population models (with an IMF extending up to 100 $M_{\star}$ and a continuous star formation in the past 30 Myr; \citealt{eldridge_binary_2017}). Following \citet{calabro_evidence_2024}, we considered the metallicity range from 0.05 to 1 times solar (i.e., 0.05, 0.1, 0.15, 0.2, 0.3, 0.4, 0.5, 0.7, 1), with the solar reference consistent with \citet{asplund_chemical_2009}. In all cases, we derived predictions for four different ionization parameters log$_{10}(\mathcal{U}) =$ $-3$, $-2.5$, $-2$, and $-1.5$ and for three gas density values ($10^{2}$, $10^{3}$, and $10^{4}\;\text{cm}^{-3}$). 

Given the low metallicity we inferred for this object, we decided to follow the approach adopted in \citet{calabro_evidence_2024} also for the dust depletion, where the metals are depleted in the beginning of our {\sc CLOUDY} calculations, and we considered that this depletion is metallicity dependent, as discussed in \citet{calabro_near-infrared_2023} (see their Table 2). 

Finally, we explored the density-bounded scenario for the nebula by varying the stopping criterion in {\sc CLOUDY} from a Lyman continuum (LyC) optical depth = 10 (fully ionization bounded case) to 0.1 (fully density-bounded case). 
This would correspond to an escape fraction of ionizing photons going from 0\% to 100\%. A density-bounded scenario typically enhances the flux of high-ionization species relative to low-ionization species. That is, this would increase the O32 and Ne3O2 indices, mimicking the effect of a high ionization parameter. However, as previously found, {\it Virgil} has a very low $f_{esc, \mathrm{LyC}}$ ($\approx 2\%$), so the increase in O32 and Ne3O2 would be very minimal ($<0.05$ dex), thus not impacting our conclusions. We refer the reader to \citet{calabro_near-infrared_2023} and \citet{calabro_evidence_2024} for a more detailed description of the models adopted in this work.

As reported by \citet{calabro_evidence_2024}, the ionization of the ISM increases with redshift, as seen in the R23$-$O32 diagram (Figure \ref{fig:ism}, left panel) from the local Universe (SDSS) to high redshifts (see redshift bins from $z\approx 4$ to $z\approx 12$ in \citealt{calabro_evidence_2024}). For comparison, we include recent high-redshift ($z \approx 4-12$) NIRSpec results from \citet{cameron_jades_2023, mascia_closing_2023, nakajima_jwst_2023, boyett_extreme_2024, calabro_evidence_2024, calabro_evolution_2024, roberts-borsani_between_2024}, along with the low-redshift sample from the Sloan Digital Sky Survey (SDSS; \citealt{york_sloan_2000, kauffmann_stellar_2003}). Additionally, we include metal-poor and high-ionization systems selected in the local Universe to approximate the properties of high-redshift galaxies (\citealt{bian_direct_2018}).

{\it Virgil} lies in the upper region of this diagram, but it is not as extreme as, for instance, GHZ2 at $z\approx 12$, analyzed by \citet{calabro_evidence_2024} and first reported by \citet{zavala_luminous_2025}. Overall, its behavior aligns with other high-$z$ galaxies observed during the Epoch of Reionization (EoR). Comparing its position with photoionization models for SFGs, {\it Virgil} overlaps with a wide range of models, from high-ionization systems (log$_{10}(\mathcal{U}) = -1.5$) with nearly solar metallicity to lower-ionization (log$_{10}(\mathcal{U}) = -2.5$) metal-poor environments, spanning a variety of electron densities ($10^2$–$10^4$ cm$^{-3}$). This degeneracy complicates its interpretation. However, independent measurements (see Section 3.2.2) suggest log$_{10}(\mathcal{U}) \approx -2$ and a metal-poor composition ($\approx 10- 20\%$ $Z_{\odot}$), help to constrain its nature. Taken together, these results suggest that {\it Virgil} shares properties with the average SFG population during EoR.

We also investigated the Ne3O2Hd$-$Ne3O2 diagram (Figure \ref{fig:ism}, right panel). As with R23$-$O32, we include SDSS galaxies from the local Universe and high-$z$ analogs from \citet{bian_direct_2018}. {\it Virgil} falls within the region occupied by other high-$z$ galaxies (e.g., \citealt{heintz_jwst-primal_2025}) and lies along the evolutionary sequence reported by \citet{calabro_evolution_2024} at $z\approx 4$–$12$. Compared to R23$-$O32, this diagram exhibits less degeneracy with photoionization models. However, model predictions in this case would suggest a lower log$_{10}(\mathcal{U})$ for metal-poor solutions, but still within the range inferred for this object when using different tracers.

\medskip

Without the prior assumption that {\it Virgil} may host an AGN, as proposed in \citet{iani_midis_2024}, these two diagrams, together with the previously derived physical properties, indicate that {\it Virgil} has a hard ionizing radiation field and ionization properties consistent with the average SFG population observed during EoR. However, its exceptionally high equivalent width for [O\,{\sc iii}]$\lambda\lambda4959,5007$ ($1514\pm18$\;\AA), combined with its low metallicity ($10-20\%\;Z_{\odot}$) and relatively low $M_{\star}$ ($\log_{10}( M_{\star}/M_{\odot}) \approx 8.5$ from \citealt{iani_midis_2024}), suggest that {\it Virgil} is a relatively low-$M_{\star}$ extreme line emitter, consistent with the sample studied by \citet{boyett_extreme_2024}, which spans similar redshifts.

Interestingly, following the discussion in \citet{rhoads_finding_2023} and the spectral properties analyzed so far, {\it Virgil} could resemble a typical Green Pea (GP; \citealt{cardamone_galaxy_2009, jaskot_origin_2013, henry_ly_2015, izotov_detection_2016, brunker_properties_2020}, among others) or a Blueberry (BB; e.g., \citealt{cameron_jades_2023, langeroodi_evolution_2023}), which are often referred to as the best nearby analogs to high-redshift galaxies.

\subsection{Does the spectrum indicate that Virgil is an AGN?}

\citet{iani_midis_2024} modeled {\it Virgil} across all available photometric bands (from HST and JWST), performing SED fitting using multiple codes, and concluded that it contains an AGN. We will re-examine this approach with improved data in Section \ref{SEDfits}. To investigate this possibility further, we exploit NIRSpec/PRISM data to search for spectral signatures indicative of an AGN.

\subsubsection{Searching for Broadening in the Balmer Lines}

\begin{figure*}
    \centering
    \includegraphics[width=0.49\linewidth]{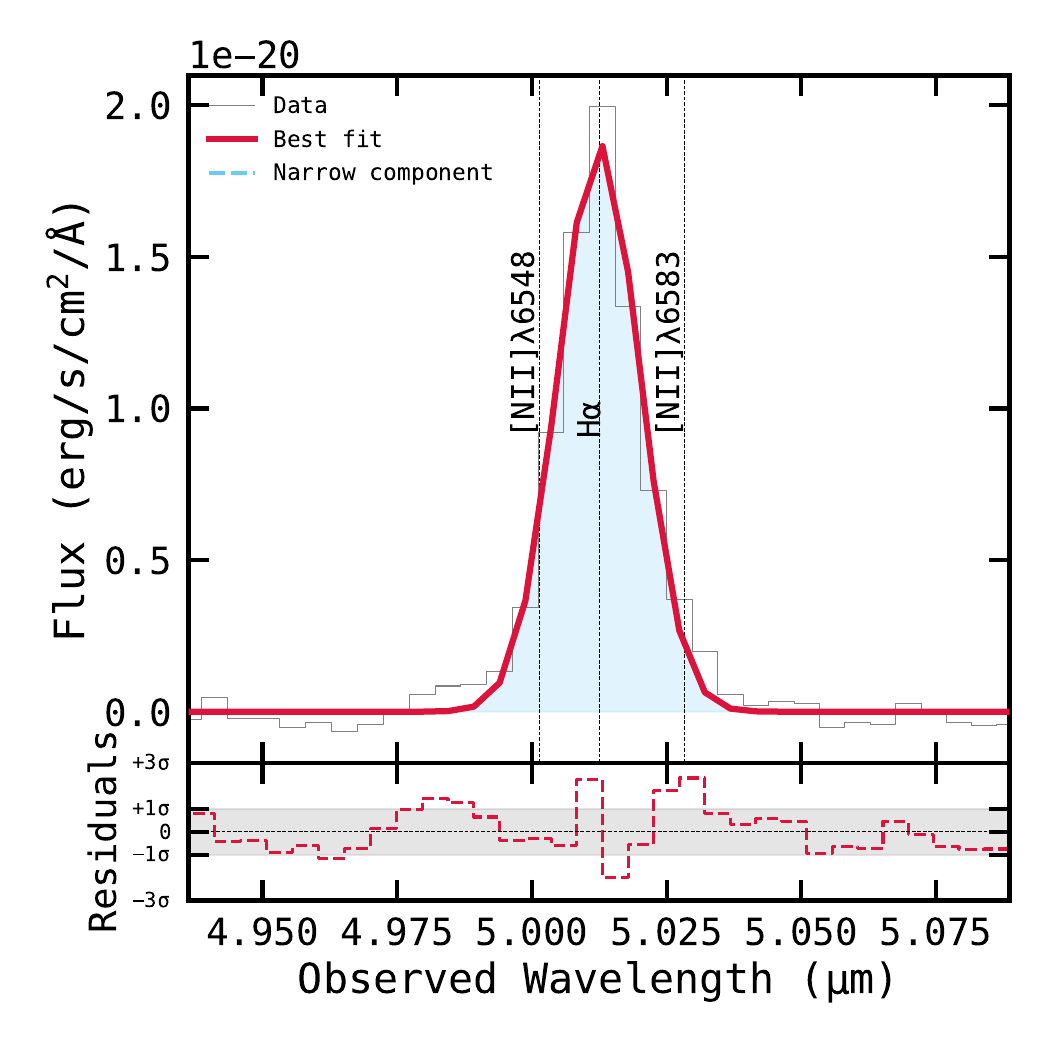}
    \includegraphics[width=0.49\linewidth]{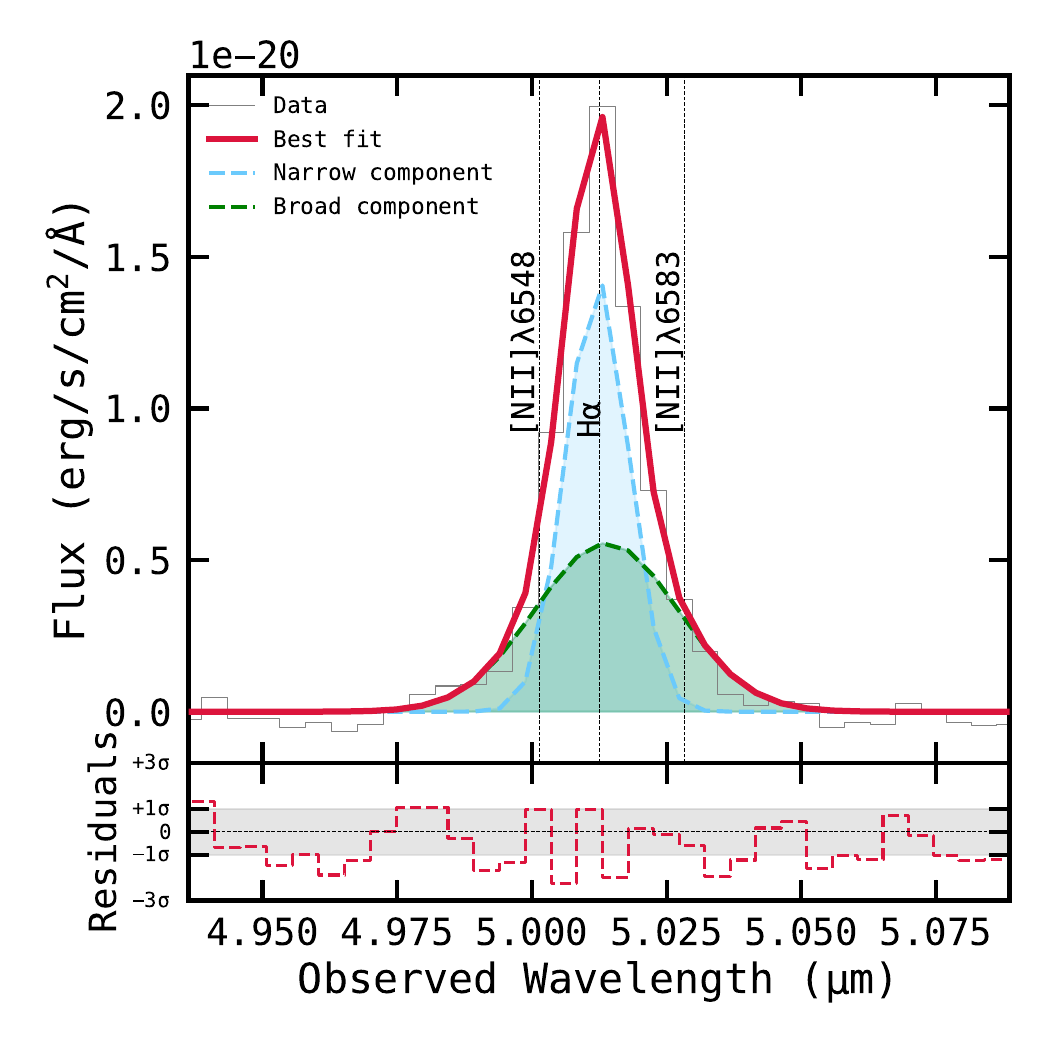}
    \caption{{\bf Left panel:} Fit of the H$\alpha$ + [NII] complex without including a broad component. {\bf Right panel:} Fit of the H$\alpha$ + [NII] complex with adding a broad component. In this case, the FWHM is $1750 \pm 575$~km~s$^{-1}$. The residuals appear to behave better when a broad component is included, although the BIC shows only a marginal improvement with the added complexity.}
    \label{fig:broad_no_broad_Ha}
\end{figure*}

\begin{figure*}
    \centering
    \includegraphics[width=1\linewidth]{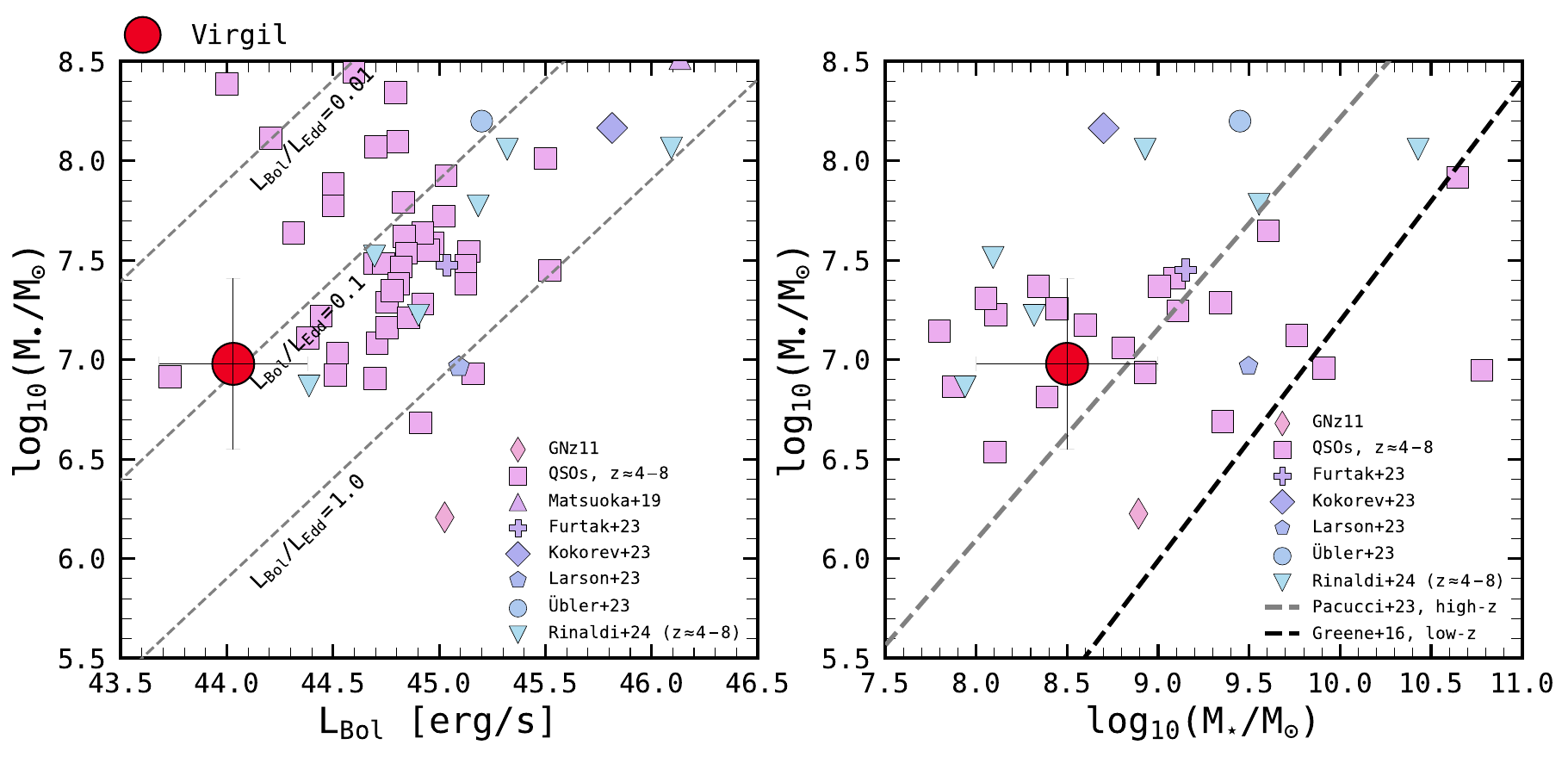}
    \caption{{\bf Left panel:} The derived $M_{\bullet}$ as a function of $L_{Bol}$, assuming {\it Virgil} has a broad component in H$\alpha$. We also show other recent findings from the literature (with some of them selected as LRDs, \citealt{rinaldi_not_2024}): \citet{matsuoka_discovery_2019, furtak_jwst_2023, harikane_jwstnirspec_2023, kocevski_hidden_2023, kokorev_uncover_2023, larson_ceers_2023, ubler_ga-nifs_2023, maiolino_small_2024}. The dashed lines represent bolometric luminosities corresponding to Eddington ratios of $L_{Bol}/L_{Edd}$ = 0.01, 0.1, and 1.0. {\bf Right panel:} The derived $M_{\bullet}$ as a function of $M_{\star}$ (stellar mass for {\it Virgil} comes from \citealt{iani_midis_2024} and its error bar reflects the overall uncertainty in fitting this object). For comparison, we show the bold black dashed line indicating the best fit to $z\approx0$ AGN samples (\citealt{greene_megamaser_2016}). The trend at higher redshifts is based on the recent analysis presented in  \citet{pacucci_jwst_2023}.}
    \label{fig:black_hole_prop}
\end{figure*}

In \citet{iani_midis_2024}, we found that {\it Virgil}'s SED resembles that of a LRD, which often exhibits broad Balmer lines (e.g., \citealt{furtak_jwst_2023, kokorev_uncover_2023, matthee_little_2024, rinaldi_not_2024}). Using \textsc{MSAEXP}, we modeled the NIRSpec/PRISM data with and without a broad component for the Balmer lines (H$\alpha$ and H$\beta$), closely examining the residuals. Figure~\ref{fig:broad_no_broad_Ha} presents the results in two panels.  To take into account the effects of the line-spread function, we additionally convolve our model with a Gaussian of variable resolution (\citealt{isobe_redshift_2023, de_graaff_ionised_2024}). On the left, we show the fit to the H$\alpha$ + [NII] complex without a broad-H$\alpha$ component, while on the right, we show the fit with the broad-H$\alpha$ component. In the latter case, the broad component is detected at the level of approximately $20\sigma$ and exhibits a full width at half maximum (FWHM) of $1750 \pm 575$ km/s. 

To assess the quality of these fits, we computed the Bayesian Information Criterion (BIC; \citealt{liddle_information_2007}) for both scenarios. The model, including the broad component, achieved the lowest BIC, indicating the best balance between goodness-of-fit and model complexity. However, the $\Delta$BIC is not large enough to definitively conclude that {\it Virgil} is a Type-1 AGN ($\Delta$BIC $<2$). We also observe a tentative detection of a broad component in H$\beta$, but its significance falls below $2\sigma$, leaving its presence uncertain.
We want to highlight that these results may be affected by the source’s position within the slit (Figure \ref{fig:virgil_spectrum}, right panel): the slit only partially covers the galaxy, with the red knot located near the edge—or possibly even in the gap between shutters—potentially limiting our sensitivity to broad-line emission.

Nonetheless, under the assumption that {\it Virgil} exhibits H$\alpha$ with a broad component, we estimated the central black hole mass ($M_{\bullet}$). Assuming the gas in the broad line region (BLR) is virialized, $M_{\bullet}$ was derived from the spectral properties of the H$\alpha$ BLR region using the calibration proposed by \citet{reines_dwarf_2013}:
\begin{align}
\log_{10}\left(\frac{M_{\mathrm{BH}}}{M_{\odot}}\right) &= \alpha + \log_{10}(\varepsilon) + \beta \log_{10}\left(\frac{L_{H\alpha, \mathrm{broad}}}{1 \times 10^{42}\; \mathrm{erg/s}}\right) \nonumber \\
&\quad + \gamma \log_{10}\left(\frac{\mathrm{FWHM_{broad}}}{1 \times 10^{3}\; \mathrm{km/s}}\right),
\end{align}
where $\alpha = 6.57$, $\beta = 0.47$, and $\gamma = 2.06$, and $\epsilon$ is the scaling factor which depends on the structure, kinematics, and orientation of the BLR. Different studies report $\epsilon$ ranging from 0.75 to 1.4 (e.g., \citealt{reines_dwarf_2013}). Following \citet{reines_relations_2015}, for our $M_{\bullet}$, we adopted  $\epsilon = 1.075 \pm 0.325$. We retrieved log$_{10}(M_{\bullet}/M_{\odot}) = 6.98\pm0.35$, which is in agreement with the estimate provided by \citet{iani_midis_2024} (based on photometry). We estimated the bolometric luminosity ($L_{bol}$) by using Equation 6 in \citet{stern_type_2012}, finding that log$_{10}(L_{bol}/(\text{erg s}^{-1})) = 44.03\pm0.43$. We also adopted Equation 1 from \citet{netzer_accretion_2009} from which we derive log$_{10}(L_{bol}/(\text{erg s}^{-1})) = 45.17\pm0.02$, which likely represents an upper limit. Also in this case, these estimates are in agreement with \citet{iani_midis_2024}. In Figure \ref{fig:black_hole_prop}  we show our measurements of $M_{\bullet}$ for {\it Virgil} in the context of the recent literature.

Based on the derived $M_{\bullet}$ from the H$\alpha$ BLR, we also derived the Eddington luminosity. We found that log$_{10}(L_{\text{Edd}}/(\text{erg s}^{-1}) = 4\pi GM_{\bullet}m_{p}c/\sigma_{t} = 45.07\pm0.35$, where $G$ is the gravitational constant, $m_{p}$ the proton mass, $c$ the speed of light, and $\sigma_{T}$ the Thomson scattering cross-section. Finally, we estimated the Eddington ratio as $\lambda_{\text{Edd}} = L_{bol}/L_{\text{Edd}}$. Depending on the choice of $L_{bol}$, the Eddington ratio ($\lambda_{\text{Edd}}$) ranges from $\approx 0.1$ to $1.23$, consistent with the typical values found for LRDs (e.g., \citealt{noboriguchi_similarity_2023}). We show our results in Figure \ref{fig:black_hole_prop} and summarize these measurements in Table \ref{tab:black_hole_prop}.

\begin{deluxetable}{lc}
\tablecaption{Black Hole Properties \label{tab:black_hole_prop}}
\tablehead{
\colhead{Property} & \colhead{Value}
}
\startdata
FWHM$^{*}$ [km/s] & $1750 \pm 575$ \\
log$_{10}(M_{\bullet}/M_{\odot})$ & $6.98 \pm 0.35$ \\
log$_{10}(L_{Bol}/(\mathrm{erg\ s}^{-1}))$$^{**}$ & $44.03 \pm 0.43 - 45.17 \pm 0.02$ \\
log$_{10}(L_{\text{Edd}}/(\mathrm{erg\ s}^{-1}))$ & $45.07 \pm 0.35$ \\
$\lambda_{\text{Edd}}$ & $0.1 - 1.23$ \\
\enddata
\tablecomments{These measurements rely on the assumption that the broadening of H$\alpha$ is real. Nonetheless, these values are in agreement with previous estimates from photometry presented in \citet{iani_midis_2024}. $^{*}$ refers to the broad component of H$\alpha$. $^{**}$ refers to the two values estimated by using both \citet{stern_type_2012} and \citet{netzer_accretion_2009} formulas.}
\end{deluxetable}

\subsubsection{Searching for AGN Signatures with Line Ratio Diagnostics}

\begin{figure*}
    \centering
    \includegraphics[width=1.\linewidth]{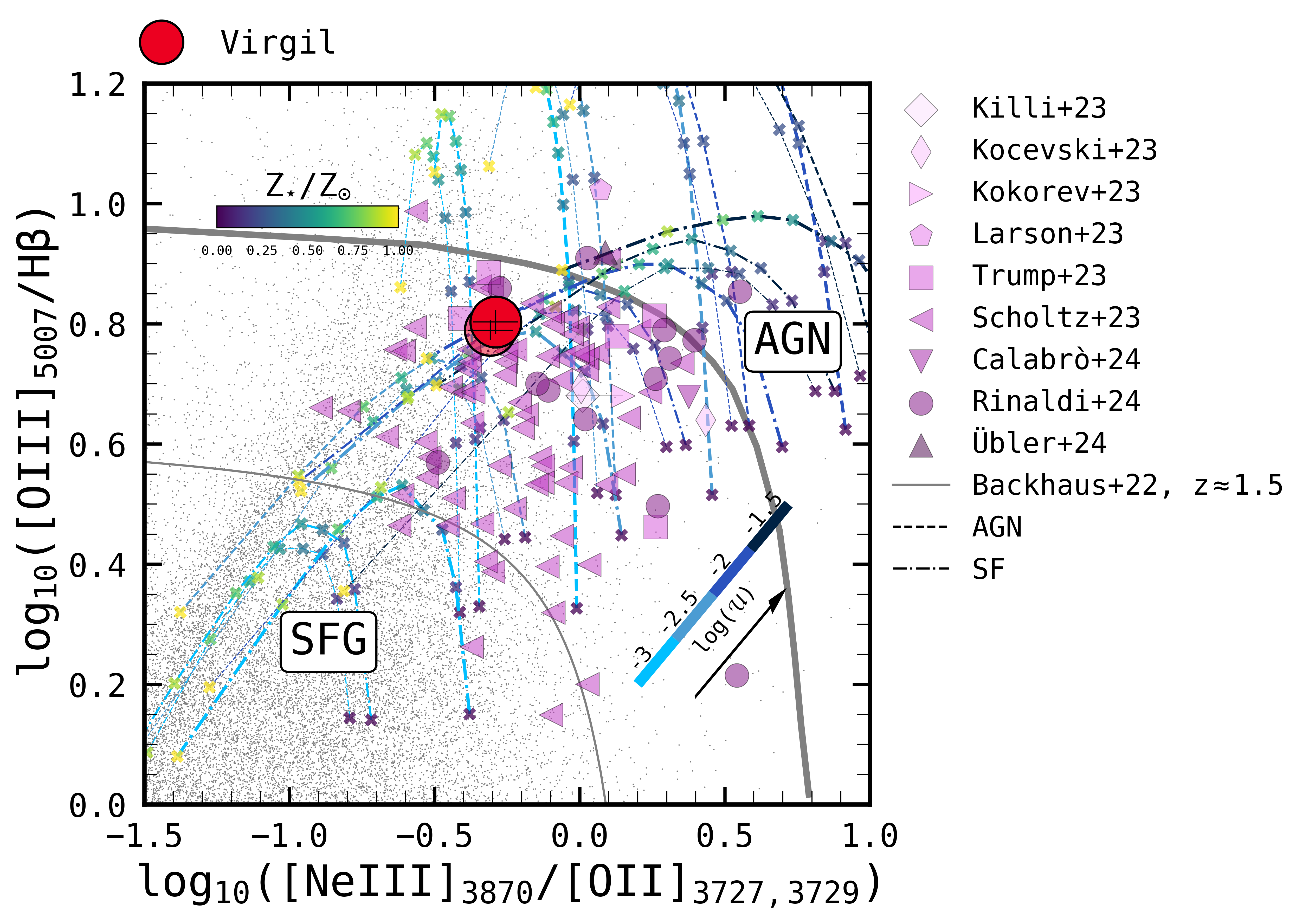}
    \caption{The “OHNO” diagram, displaying the following ratios: [OIII]$\lambda$5007/H$\beta$ vs. [NeIII]$\lambda$3870/[OII]$3727,3728$. {\it Virgil} is shown as a red filled circle, with higher opacity representing the case without dust correction and lower opacity indicating the dust-corrected one. We also show the recent literature at $z\gtrsim4$ from \citet{killi_deciphering_2023, kocevski_hidden_2023, kokorev_uncover_2023, larson_ceers_2023, trump_physical_2023, ubler_ga-nifs_2024, calabro_evidence_2024,  rinaldi_not_2024}.
    In particular, sources from \citet{rinaldi_not_2024} have been photometrically selected as LRDs, with some of them clearly showing broad components in their Balmer lines from NIRSpec data. For comparison, we also show SDSS (SFGs and AGNs) at low-z ($z\approx0$) from \citet{york_sloan_2000, kauffmann_stellar_2003}. The separation line comes from \citet{backhaus_clear_2022} at $z\approx1.5$. We also show an extrapolation at $z\approx6.5$ (increased thickness), based on \citet{ backhaus_ceers_2024}. For comparison, theoretical predictions (for both SFGs and AGNs; dashdot and dashed respectively) are shown with the following settings: $n_{e}$ (10$^{2}$, 10$^{3}$, and 10$^{4}$ cm$^{-3}$; increasing thickness), log$_{10}(\mathcal{U})$ (-3, -2.5, -2, -1.5; different shades of blue), and different metallicities (0.05 to solar; color-coded in viridis using “x” symbols).}
    \label{fig:ohno}
\end{figure*}

\begin{figure*}
    \centering
    \includegraphics[width=1\linewidth]{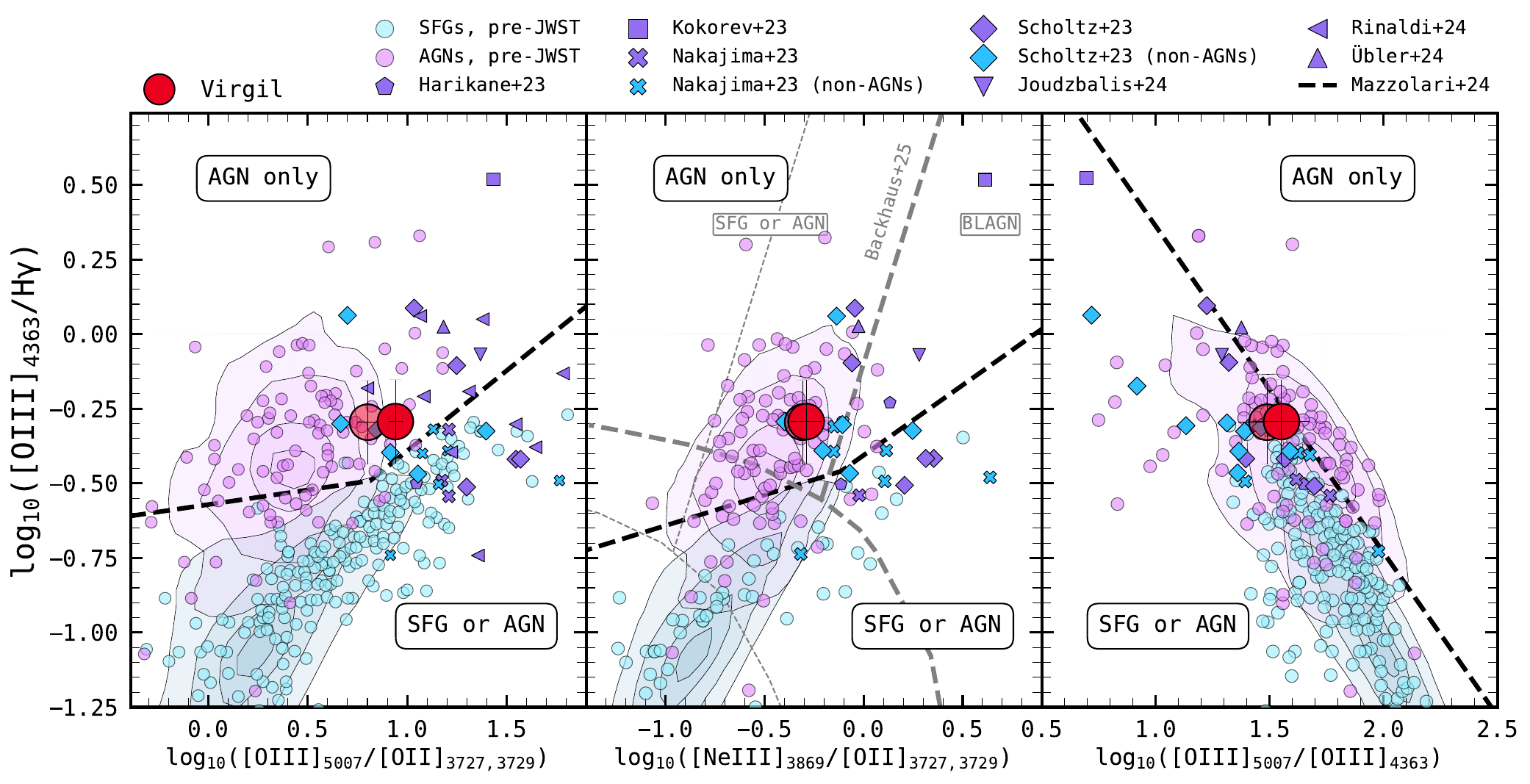}
    \caption{{\bf Left Panel:} O32 vs. O3Hg. {\bf Middle Panel:} Ne3O2 vs. O3Hg. {\bf Right Panel:} O33 vs. O3Hg. {\it Virgil} is shown as a red filled circle, with higher opacity representing the case without dust correction and lower opacity indicating the dust-corrected one. The dashed black lines indicate the separation criteria proposed by \citet{mazzolari_new_2024} to distinguish AGNs from SFGs. The contour areas correspond to the SDSS sample in the local Universe (\citealt{abazajian_seventh_2009}) for comparison. Filled circles represent pre-JWST literature data for SFGs and AGNs  (mostly at low redshift; \citealt{seyfert_nuclear_1943, izotov_chemical_2006, berg_direct_2012,  amorin_extreme_2015, perna_x-raysdss_2017, yang_blueberry_2017, yang_ly_2017, izotov_j08114730_2018, dors_chemical_2020, armah_chemical_2021,  pustilnik_xmp_2021}). For comparison with recent JWST-based studies, we include results from \citet{harikane_jwstnirspec_2023, nakajima_jwst_2023, scholtz_jades_2023, ubler_ga-nifs_2024, juodzbalis_dormant_2024, rinaldi_not_2024} (with some of them having both SFGs and AGNs). In particular, the sample from \citet{rinaldi_not_2024} consists of photometrically selected LRDs in GOODS-S, some exhibiting clear broadening in the Balmer lines. The gray dashed line in the middle panel represents the separation criterion at $z\gtrsim4$ recently proposed by \citet{backhaus_emission-line_2025} to distinguish BLAGNs from non-BLAGNs.
}
    \label{fig:virgil_line_ratios_oiii_hg}
\end{figure*}

Disentangling SFGs from AGNs at high redshift is particularly challenging. Classical diagnostics, such as the Baldwin, Phillips \& Terlevich (BPT; \citealt{baldwin_classification_1981}) diagram, were calibrated for low-redshift galaxies and strongly depend on metallicity, making them less reliable in the early Universe. At high redshift, galaxies are typically more metal-poor \citep[e.g.,][]{topping_metal-poor_2024}, with younger stellar populations and higher ionization parameters. These conditions shift SFGs toward the AGN locus in BPT-like diagrams, while AGN Narrow Line Regions (NLRs) often overlap with the SFG sequence. As a result, traditional line-ratio diagnostics become increasingly ambiguous, as also predicted by theoretical models \citep{nakajima_diagnostics_2022}.

Although we are not able to resolve [N\,\textsc{ii}]$\lambda6548$ from H$\alpha$ and both [O\,\textsc{i}]$\lambda6302$ and the [S\,\textsc{ii}]$\lambda\lambda6717,6731$ doublet have SNR $\lesssim 2$, it is important to stress that classical BPT-like diagrams may not reliably determine whether \textit{Virgil} hosts an AGN. Even when these lines are available, recent studies have shown that AGNs can occupy the same region as SFGs in BPT diagrams. For example, the $z \approx 5.55$ galaxy GS\_3073 shows clear AGN signatures through broad H$\alpha$ and H$\beta$ emission, yet it lies within the SFG region in the BPT diagram \citep{ubler_ga-nifs_2023}. Like {\it Virgil}, GS\_3073 is extremely metal-poor (12 + log(O/H) $=$ 8.02), underscoring how low metallicity can blur the distinction between star formation- and AGN-dominated galaxies at high redshift. Similar conclusions have been reported by \citet{scholtz_jades_2023} and \citet{kocevski_hidden_2023}, in agreement with photoionization models from \citet{nakajima_diagnostics_2022}, which predict significant overlap between low-metallicity SFGs and AGNs in classical diagnostics.

Since the launch of JWST, the limitations of classical diagnostics have led to renewed efforts to develop alternative line-ratio diagrams optimized for high-redshift galaxies (e.g., \citealt{hirschmann_emission-line_2023, mazzolari_new_2024, shapley_aurora_2024, backhaus_emission-line_2025}). In the following, we employ a set of diagnostic diagrams that have been extensively used to distinguish AGNs from SFGs at high redshifts.

\paragraph{{\bf The “OHNO” diagram}}

One widely used high-redshift diagnostic is the “OHNO” diagram (\citealt{trouille_pushing_2011, zeimann_hubble_2015, backhaus_clear_2022, backhaus_clear_2023, cleri_using_2023, trump_physical_2023, feuillet_classifying_2024}), which is based on Ne3O2 and R3 and has proven to be a robust ionization diagnostic for high-redshift galaxies. Ne3O2 is particularly useful as it involves emission lines with similar ionization energies and closely spaced wavelengths, minimizing dust attenuation effects. 
As with the R23–O32 and Ne3O2Hd–Ne3O2 diagrams, we also included photoionization models, following the configuration adopted by \citet{calabro_evidence_2024}. For AGN models, we adopted the default {\sc CLOUDY} prescription, assuming a multi-component power-law continuum with a “blue bump” temperature of $10^{6}$ K and spectral energy indices of $\alpha_{\rm UV} = -0.5$, $\alpha_{\rm X} = -1.35$, and $\alpha_{\rm OX} = -1.4$ (\citealt{groves_dusty_2004}) for the UV, X-ray, and optical-to-X-ray ranges, respectively. We used the same grid of metallicities (0.05 to 1 solar) as for the models for SFGs. We refer the reader to \citet{calabro_near-infrared_2023} and \citet{calabro_evidence_2024} for a more detailed discussion of these models.

It is worth noting that the OHNO diagram may be affected by its dependence on metallicity (\citealt{tripodi_spatially_2024}), where metal-poor objects, such as {\it Virgil}, can appear as outliers (\citealt{scholtz_jades_2023}).

In Figure \ref{fig:ohno}, we show the “OHNO” diagram with {\it Virgil} as well as the recent literature on other high-$z$ sources (\citealt{killi_deciphering_2023, kocevski_hidden_2023, kokorev_uncover_2023, larson_ceers_2023, trump_physical_2023, scholtz_jades_2023, ubler_ga-nifs_2024, calabro_evidence_2024, rinaldi_not_2024}; with some of them identified as Type 1 and Type 2 AGNs) and, for comparison, the sample of SFGs and AGNs at $z\approx0$ from \citet{york_sloan_2000, kauffmann_stellar_2003}.  

{\it Virgil} exhibits moderately high Ne3O2 and R3 ratios\footnote{We show our measurements before and after applying dust corrections.} and it lies well above the separation line proposed by \citet{backhaus_clear_2022} at $z\approx1.5$. Thus, it is consistent with AGN model predictions at subsolar metallicities ($Z_{\rm gas}/Z_{\odot} \lesssim 0.2$), moderate ionization parameters ($\log_{10}(\mathcal{U}) \approx -3$), and $n_{e} \approx 10^{3}-10^{4}\;\text{cm}^{-3}$. However, {\it Virgil} is also consistent with SF models, overlapping with the parameter space expected for H\,{\sc ii} regions with high $n_{e}$ ($\approx 10^{3}-10^{4}\;\text{cm}^{-3}$), high ionization parameter ($\log_{10}(\mathcal{U}) \approx -2$), and wide range of metallicities.

Interestingly, {\it Virgil} occupies a region of the parameter space consistently populated by LRDs (some of which exhibit clear broadening in their Balmer lines; \citealt{rinaldi_not_2024}), as well as other recently discovered AGNs (including Type 2; e.g., \citealt{kocevski_hidden_2023, kokorev_uncover_2023, larson_ceers_2023, scholtz_jades_2023, ubler_ga-nifs_2024}). However, as shown in \citet{calabro_evidence_2024} and supported by {\sc CLOUDY} model predictions, this region of parameter space is also consistently populated by SFGs with a different set of properties.

We then utilized the recent results of \citet{backhaus_clear_2022, backhaus_emission-line_2025} to extrapolate the separation line up to $z\approx6.5$ (thick gray line), relying on the proposed redshift evolution of R3 and N3O2. In this scenario, {\it Virgil}, along with sources from \citet{killi_deciphering_2023}, \citet{scholtz_jades_2023}, \citet{trump_physical_2023}, as well as LRDs from \citet{rinaldi_not_2024}, and the BL AGN from \citet{kokorev_census_2024}, would fall below the extrapolated separation line, placing {\it Virgil} within the SFG region—or at least blurring the net distinction that would otherwise arise using the relation at $z\approx1.5$. 

We emphasize that this is only a test to explore the implications of shifting the separation line to higher redshifts, based on the evolutionary trends proposed by \citet{backhaus_clear_2022, backhaus_emission-line_2025}, and should not be interpreted as a definitive redefinition. Nonetheless, applying diagnostic diagrams calibrated at lower redshifts complicates the classification of {\it Virgil} and could provide misleading results in general.

Considering the degeneracies revealed by {\sc CLOUDY} models in the region of the parameter space occupied by {\it Virgil} and the possible redshift evolution of R3 and Ne3O2, both the SFG and AGN scenarios (or a mixed nature) remain plausible for this source.

\paragraph{{\bf The [O\;{\sc iii}]$\lambda$4363 diagrams}}
\citet{mazzolari_new_2024} proposed new separation lines for three diagnostic diagrams based on [O\;{\sc iii}]$\lambda$4363, which has been linked to AGN activity (\citealt{brinchmann_high-z_2023}) with higher ISM temperatures driven by AGN ionizing radiation (\citealt{ubler_ga-nifs_2024}): 

\begin{enumerate}
    \item O32 vs. O3Hg
    \item Ne3O2 vs. O3Hg
    \item O33 vs. O3Hg
\end{enumerate}

From a theoretical point of view, enhanced [O\;{\sc iii}]$\lambda$4363 emission could hint at the presence of an AGN, as the ionizing photons produced by an AGN are typically more energetic than those from star formation, thus leading to significantly more efficient gas heating. Notably, it has been proposed that strong [O\;{\sc iii}]$\lambda$4363-emitting regions may coincide with high-ionization nuclear emission-line regions (HINERs; \citealt{binette_photoionization_1985}), and that an enhanced [O\;{\sc iii}]$\lambda$4363/[O\;{\sc iii}]$\lambda$5007 ratio ($R_{[OIII]}$) would imply a strong connection with AGN activity. For {\it Virgil} we find that $R_{[OIII]} =0.032\pm0.01$, which is similar to what has been observed in other Seyfert 2s: ESO 138-G01, Mrk 1210, and NGC 4507, for which \citet{binette_constraints_2024} concluded that a high-density NLR component is likely present. In particular, the combination of such high $R_{[OIII]}$ and relatively high R3 would imply that their plasma is much hotter ($\approx18000-20000$ K), possibly as a result of fast shocks (e.g., \citealt{binette_radiative_1985}). Following this line, \citet{nagao_where_2001} found, for a sample of low-$z$ objects, that a high $R_{[OIII]}$ ratio correlates with objects exhibiting hotter MIR colors.

\medskip

Below, we briefly overview each diagnostic diagram (already presented in \citealt{mazzolari_new_2024}). 

We want to highlight that dust reddening can significantly affect the line ratios, particularly for the O32$-$O3Hg diagram, due to the wavelength separation between the adopted lines. This effect can cause AGNs to shift toward the SFG+AGN locus when there is substantial reddening. Nonetheless, the inverse scenario does not hold, as SFGs would not populate the AGN locus under any circumstances. The Ne3O2$-$O3Hg diagram, instead, is very similar to the “OHNO” diagram, but it is more stable as the separation between SFG and AGN is mainly based on a different gas temperature.

For O32$-$O3Hg, \citet{mazzolari_new_2024} showed that high-$z$ sources are generally distributed toward the upper region of the diagram compared to local samples (SDSS, contour area). Moreover, they point out that a non-negligible fraction of high-$z$ sources not classified as AGNs falls within the region occupied by local AGNs. In contrast, some high-$z$ AGNs overlap with the area covered by their local analogs. Following the discussion in \citet{mazzolari_new_2024} and their photoionization models, SFGs should occupy a well-constrained region of the O32$-$O3Hg diagram, with their upper boundary closely matching the distribution of SDSS SFGs and local analogs. In contrast, AGNs are expected to populate both the SFG region and the area occupied by local AGN samples. They show that SFG models do not extend into the AGN-dominated region. However, AGN models can reach the SFG region for some local and high-$z$ AGNs. They also conclude that these trends remain unchanged even when considering the entire grid of SFG models from \citet{feltre_nuclear_2016}, reinforcing the idea that AGN-driven photoionization can produce significantly higher [O\,{\sc iii}]$\lambda$4363/H$\gamma$ ratios than ionization from hot stars. This is likely due to the higher-energy ionizing photons in AGNs, which heat the gas more efficiently at a given ionization parameter.

As with the O32$-$O3Hg diagram, normal SFGs and local analogs primarily occupy the lower region of the Ne3O2$-$O3Hg parameter space, while AGNs are now distributed across a broader area, including the upper-left region, which is not reached by any of the SFG models that \citet{mazzolari_new_2024} tested or by observed SFG samples. Notably, they also find that sources classified as AGNs in the O32$-$O3Hg diagram remain above the SFG distribution in this diagnostic. Likewise, sources not explicitly identified as AGNs but found within the AGN region of the first diagnostic continue to occupy the same region in the Ne3O2$-$O3Hg diagram.

The third diagnostic compares O33 and O3Hg. The overlap between local SFGs and AGNs is significantly larger than in the previous diagnostics. However, a distinct region characterized by high [O\,{\sc iii}]$\lambda$4363/H$\gamma$ and high [O\,{\sc iii}]$\lambda$5007/[O\,{\sc iii}]$\lambda$4363 remains populated exclusively by local AGNs. As noted by \citet{mazzolari_new_2024}, very few high-$z$ galaxies fall within the AGN-only region, with most shifting toward lower [O\,{\sc iii}]$\lambda$5007/[O\,{\sc iii}]$\lambda$4363 values. They also show that current photoionization models do not cover this region.

According to \citet{mazzolari_new_2024}, the distribution of SFG models closely follows that of local galaxies, particularly at the boundary with the AGN-only region. In contrast, AGN models extend across both the SFG and AGN regions. They also noted that when considering the full parameter grid of \citet{feltre_nuclear_2016}, SFG and AGN models would overlap entirely.

\medskip

In the case of {\it Virgil}\footnote{For {\it Virgil}, [O\;{\sc iii}]$\lambda$4363 and H$\gamma$ were successfully deblended, with both lines detected at SNR $>5$.}, we find that it consistently lies above or on the separation lines proposed by \citet{mazzolari_new_2024} (Figure \ref{fig:virgil_line_ratios_oiii_hg}), overlapping with AGNs and LRDs identified in recent studies based on JWST \citep[e.g.,][]{ kokorev_uncover_2023,
nakajima_jwst_2023,
scholtz_jades_2023, ubler_ga-nifs_2023, juodzbalis_dormant_2024, rinaldi_not_2024}. 

Nonetheless, one should note that [O\,{\sc iii}]$\lambda$4363 could be contaminated by [Fe\,{\sc ii}]$\lambda$4360. However, \citet{curti_new_2017} showed that such contamination primarily occurs in galaxies with an enriched ISM, whereas {\it Virgil} is metal-poor. Additionally, \citet{mazzolari_new_2024}, by using {\sc CLOUDY} models over a wide range of metallicities ($-2 \leq \text{log}(Z/Z_{\odot}) \leq 0.5$) and ionization parameters ($-4 \leq \text{log}_{10}(\mathcal{U}) \leq -1$), found that the [Fe\,{\sc ii}]$\lambda$4288/[Fe\,{\sc ii}]$\lambda$4360 ratio is roughly constant ($\approx 1.25$). Therefore, they searched for [Fe\,{\sc ii}]$\lambda$4288 in their sample populating the “AGN-only” region and found no evidence of its presence, concluding that any possible contamination from [Fe\,{\sc ii}]$\lambda$4360 can be considered negligible.

As with the “OHNO” diagram, we also explored the potential redshift evolution of key line ratio diagnostics. Specifically, we examined the Ne3O2 versus O3Hg plane (Figure~\ref{fig:virgil_line_ratios_oiii_hg}, middle panel), based on the empirical evolution recently proposed by \citet{backhaus_emission-line_2025}. We display their separation lines at $z \approx 0$ and $z \approx 6.5$ (gray lines, with increased thickness). Under this redshift evolution, \textit{Virgil} would fall in the SFG/AGN region, suggesting a mixed nature. 

We want to highlight that \citet{backhaus_emission-line_2025}, based on photoionization models, found that only AGN-dominated sources populate the region above the separation line proposed by \citet{mazzolari_new_2024} in this diagram. Indeed, taken at face value, these line ratio diagnostics would clearly place \textit{Virgil} in the AGN regime along with other claimed AGNs at similar redshifts. A similar argument can be made for the O32 and O33 diagnostics shown in the other two panels.

Altogether, this again underscores the complexity of distinguishing AGNs from SFGs at high redshift—especially when accounting for redshift evolution—as illustrated by \textit{Virgil}’s position in a region indicative of a mixed nature.

\medskip

Being in the ``AGN'' zone in the adopted diagnostic diagrams is a {\it condicio sine qua non} for being an AGN, as shown in Figures \ref{fig:ohno} and \ref{fig:virgil_line_ratios_oiii_hg}, although some galaxies in these regions have substantial star formation. If interpreted with no redshift correction, the two sets of diagnostic diagrams analyzed in this section would indicate that {\it Virgil} may host an AGN: (1) its position in the “OHNO” diagram aligns with regions typically occupied by AGNs and LRDs; while (2) in the [O\,{\sc iii}]$\lambda$4363-based diagrams from \citet{mazzolari_new_2024}, {\it Virgil} occupies a region predominantly populated by recently discovered AGNs at high-$z$. However, this classification strongly depends on the assumption that the criteria to separate SFGs and AGNs remain unchanged with redshift. If the redshift evolution of the relevant line ratios is considered—particularly in light of recent findings \citep[e.g.,][]{backhaus_emission-line_2025}—the classification of {\it Virgil} becomes more ambiguous, potentially suggesting a mixed nature or even a galaxy dominated by star formation.

In summary, we have conducted a detailed analysis of \textit{Virgil} based on its emission-line spectrum. At first glance, {\it Virgil} appears to be a typical galaxy for its redshift, with low-to-moderate attenuation in its emission lines and low but representative metallicity. While its emission-line ratios could suggest AGN activity, this classification becomes ambiguous when accounting for the redshift evolution of diagnostic diagrams. Had it not been identified through its pronounced spectral steepening in the MIRI bands (\citealt{iani_midis_2024}), \textit{Virgil} could easily be misclassified as an ordinary high-$z$ SFG, sharing many properties with the bulk of the galaxy population during the EoR.

\section{Photometry and Spectral Energy Distribution of Virgil}
\label{SEDfits}

\begin{figure}
    \centering
    \includegraphics[width=0.9\linewidth]{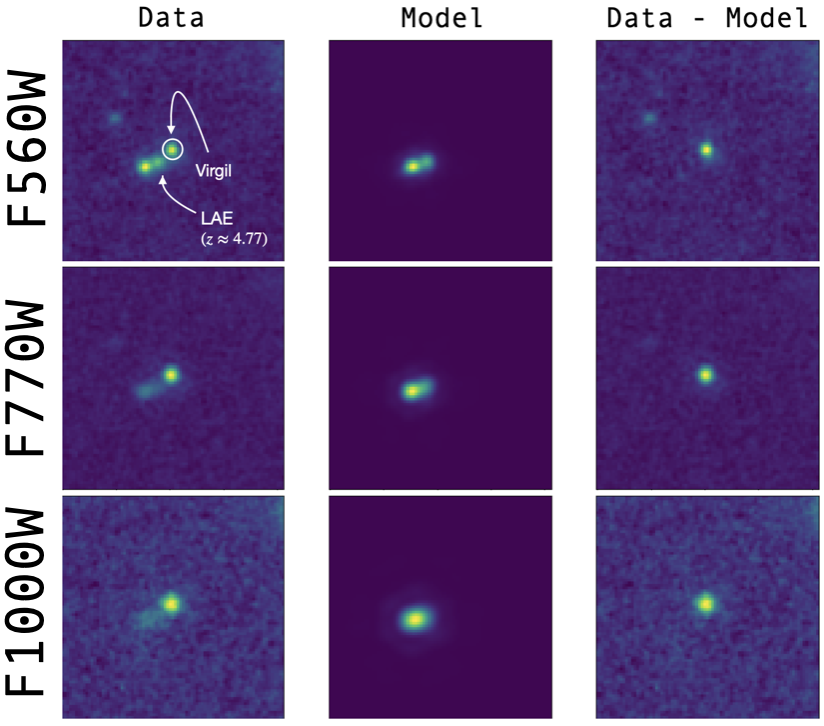}
    \caption{Illustrative example of the modeling and subtraction procedure applied to the foreground LAE at $z \approx 4.77$ located near \textit{Virgil}. Each row corresponds to a different MIRI filter: F560W (top), F770W (middle), and F1000W (bottom). The three columns show, from left to right, the original data, the best-fit model of the LAE, and the residual after model subtraction ({\tt data}$–${\tt model}). No modeling was applied to \textit{Virgil}, which is visible in all panels and highlighted in the F560W stamp. Note that the displayed scales are optimized independently for each panel for visual clarity and are not uniform across filters.
}
    \label{fig:miri_model}
\end{figure}

\begin{figure}
    \centering
    \includegraphics[width=0.9\linewidth]{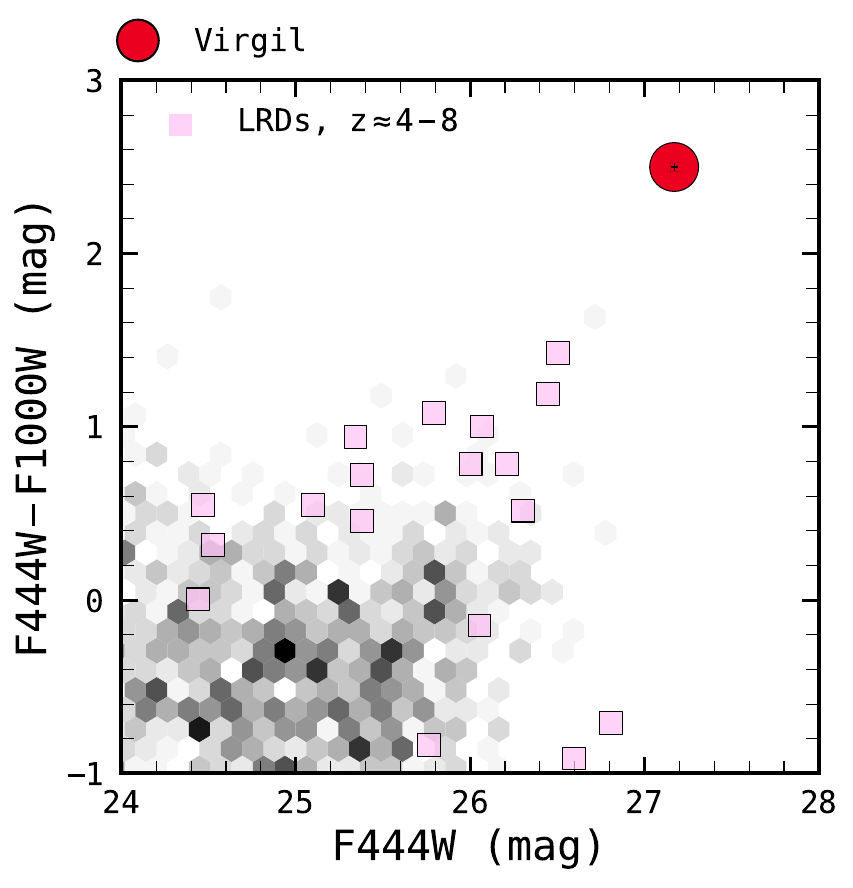}
    \caption{The color-magnitude diagram shows F444W along x-axis and F444W$-$F1000W along y-axis, including \textit{Virgil}, data from SMILES and JADES (hexagons), and a sample of LRDs in GOODS-S at $z\approx4$--8 previously reported by \citet{kokorev_census_2024}, \citet{perez-gonzalez_nircam-dark_2024},
 and \citet{rinaldi_not_2024} with MIRI detection in the SMILES catalog in GOODS-S.}
    \label{fig:virgil_color_mag}
\end{figure}

The nature of {\it Virgil} remains ambiguous when relying solely on data below 5~$\mu$m, even when rest-frame optical emission lines are considered, but changes significantly with the addition of longer wavelength observations. We leveraged these observations to perform updated photometry and SED fitting since deeper MIRI data are now available at 7.7, 10, and 15 $\mu$m from the MIDIS and PAHSPECS teams. We build on the SED fitting presented in \citet{iani_midis_2024}, now incorporating the latest MIRI data. As the derived physical properties remain essentially unchanged, we refer the reader to \citet{iani_midis_2024} for a detailed description of the modeling and results. Here, we focus on a more interpretative discussion in light of the spectral analysis presented in this work.

\subsection{Forward Modeling and Photometry of {\it Virgil}}
\begin{figure*}
    \centering    \includegraphics[width=0.96\linewidth]{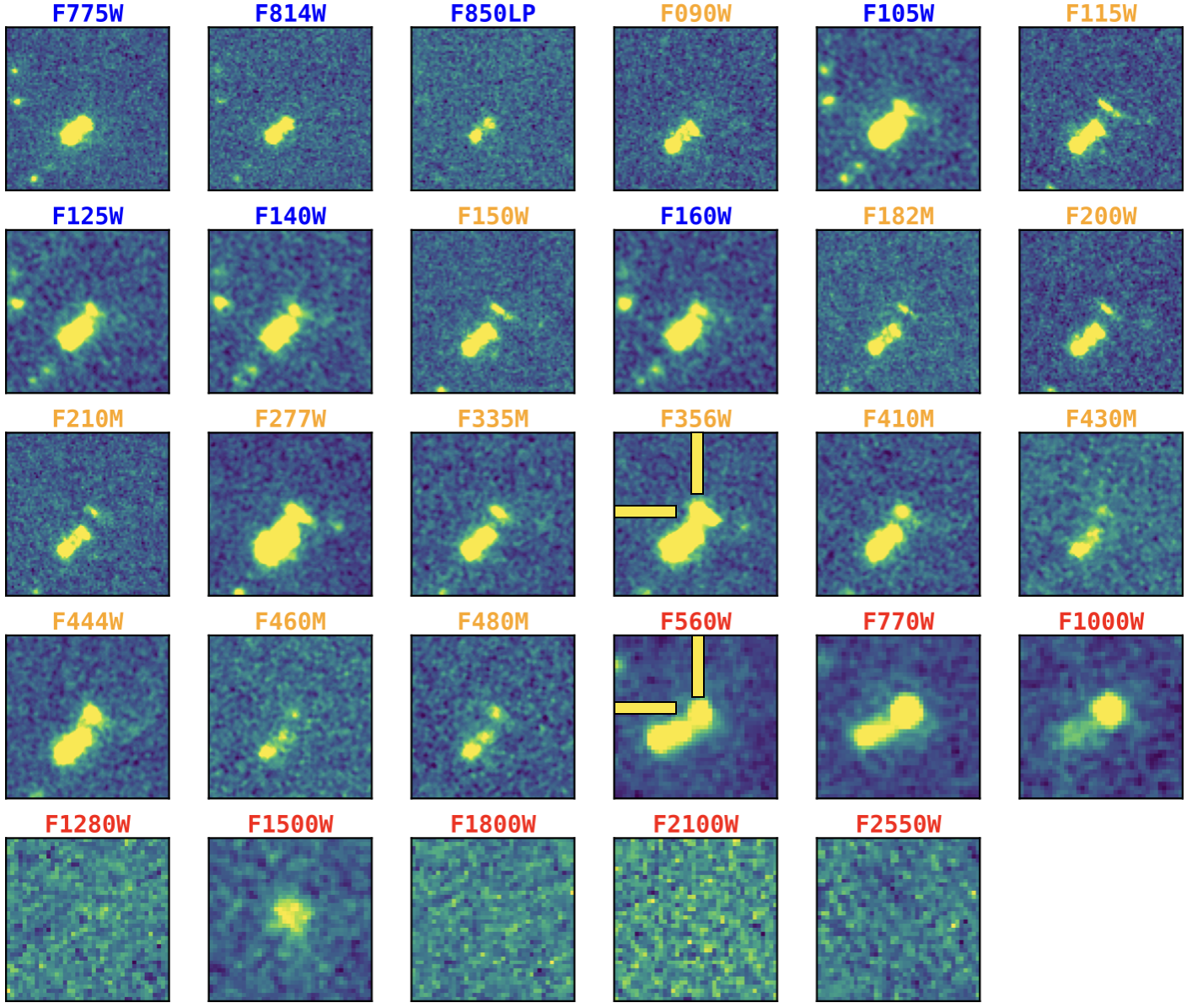}
    \caption{Cutouts ($2.5\arcsec \times 2.5\arcsec$) centered on {\it Virgil}, showing imaging from HST (from F775W only), NIRCam, and MIRI, each displayed at their native pixel scales (30 mas for HST and NIRCam, 60 mas for MIRI), and color-coded in blue, orange, and red, respectively. {\it Virgil} is highlighted in F356W and F560W. Its elongated morphology, featuring two distinct knots, is clearly visible in the NIRCam bands thanks to their superior resolution. F560W, F770W, and F1000W are from MIDIS; F1500W is from PAHSPECS, while the remaining MIRI bands are from SMILES. Each postage stamp is displayed using its intensity scale centered around the median pixel value to enhance visual contrast across filters.}
    \label{fig:virgil_postage}
\end{figure*}

We performed a set of customized photometric measurements using the {\sc photutils} package (\citealt{bradley_astropyphotutils_2022}). Object centroids were computed using the “windowed positions” method implemented in Source Extractor (\citealt{bertin_sextractor_1996}). Depending on the filter, we employed a combination of Kron apertures (\citealt{kron_photometry_1980}) with a Kron parameter of 2.5 and circular apertures. This approach was crucial as {\it Virgil} appears elongated, particularly at shorter wavelengths (NIRCam short channel), as noted in \citet{iani_midis_2024}. In Figure \ref{fig:virgil_color_mag}, we show Virgil's color F444W$-$F1000W ($2.50\pm0.02$~mag) with respect to the SMILES sample in GOODS-S as well as a sample of LRDs at $z\approx4-8$ already reported in \citet{kokorev_census_2024}, \citet{perez-gonzalez_nircam-dark_2024}, and \citet{rinaldi_not_2024}. Interestingly, at 15 $\mu$m, {\it Virgil} shows F444W$-$F1500W = $2.87\pm0.04$~mag, which highlights how extreme this source is at the MIRI wavelengths. An overview of {\it Virgil}'s appearance is presented in Figure~\ref{fig:virgil_postage}.

For circular aperture photometry, we adopted filter-dependent radii, with apertures as large as 0.4$-$0.5\arcsec\; in the MIRI bands. In each case, we measured the background in a close region around {\it Virgil} and calculated the noise using nonadjacent pixels (5 pixels apart) to take into account noise correlation, as explained in \citet{perez-gonzalez_life_2023}. We corrected the photometry to total flux using aperture corrections from JADES (for HST and NIRCam) and SMILES (for MIRI). We validated our new measurements by comparing them with those reported in \citet{iani_midis_2024}, finding good agreement.

We used \textsc{Forcepho} to model and subtract the nearby LAE ($z_{spec}\approx4.77$) to avoid light contamination during the photometric measurements. The LAE and {\it Virgil} were modeled using \citet{sersic_atlas_1968} profiles, with one Sérsic profile representing the extended flux of the LAE and four additional compact Sérsic profiles modeling its stellar clumps. We observed that {\it Virgil} consists of two knots, one of which likely drives the strong upturn in the SED at MIRI wavelengths, leading us to model these knots separately with two Sérsic profiles. We utilized \textsc{Forcepho} to sample all model parameters using NIRCam and HST images, employing the Hamiltonian Monte Carlo Markov Chain method. Each galaxy was then forward-modeled in the MIRI images by fitting the flux of each component using non-negative least squares regression, incorporating effective point-spread-function (ePSF) models provided by \citet{libralato_high-precision_2024}. Figure \ref{fig:miri_model} presents the LAE-subtracted image using the best-estimated model, with minimal residuals at the LAE’s location, demonstrating the quality of our modeling. Although we obtained a model flux of {\it Virgil} during the process, we do not adopt it as fiducial, as the complex morphology of {\it Virgil} is not well represented by two Sérsic profiles. Nevertheless, we find good agreement between the model flux and the photometry results from the LAE-subtracted images.

\subsection{Virgil is a Little Red Dot}

First, we examined the photometric criteria for LRDs. \citet{iani_midis_2024} demonstrated that {\it Virgil} satisfies most of the commonly used color criteria for LRDs, except for a relatively blue  F277W$-$F444W (\citealt{kokorev_census_2024, rinaldi_not_2024}), which explains why it was not selected in recent LRD studies in GOODS-S (\citealt{kokorev_census_2024, perez-gonzalez_what_2024, rinaldi_not_2024}). To further characterize {\it Virgil}, we adopted the methodology of \citet{kocevski_rise_2024} and estimated the continuum slope ($f_{\lambda} = \lambda^{\beta}$) between two continuum bands to identify a V-shaped SED. Specifically, following \citet{lin_discovery_2024}, we applied the criteria $\beta_{UV} < -0.37$ and $\beta_{opt} > 0$, where $\beta = -0.4((m_{1} - m_{2})/\text{log}_{10}(\lambda_{1}/\lambda_{2})) - 2$. Using F430M and F444W, we obtained $\beta_{opt} \approx 0.87$, while for $\beta_{UV}$, we found $\beta_{UV} \approx -0.68$ by relying on F115W and F150W. This places {\it Virgil} within the locus of V-shaped SED galaxies,  reinforcing its classification as a LRD. These findings underscore the complexity of LRD selection (\citealt{hainline_investigation_2024}) and suggest that refining color criteria, potentially incorporating MIRI photometry when available, may be necessary to select {\it Virgil}-like sources during EoR. Finally, as highlighted in \citet{iani_midis_2024}, we found that {\it Virgil} satisfies the compactness criterion, showing $c_{\text{F444W}}(\equiv F(0.5$\arcsec$)/F(0.25$\arcsec$))\approx1.5$.

 \section{Stellar properties of {\it Virgil}}

\subsection{SED fitting }
We followed the approach adopted in \citet{iani_midis_2024} and employed multiple SED-fitting codes to estimate the stellar properties of {\it Virgil}. While \citet{iani_midis_2024} present a detailed SED fitting analysis, we focused on deriving key stellar parameters to contextualize our spectral results and investigate the nature of this object. Specifically, we use {\sc Bagpipes} \citep{carnall_vandels_2019}, {\sc CIGALE} \citep{boquien_cigale_2019}, {\sc Prospector} \citep{johnson_stellar_2021}, and {\sc Synthesizer-AGN} \citep{perez-gonzalez_stellar_2003, perez-gonzalez_stellar_2008, perez-gonzalez_what_2024}. As demonstrated and largely discussed in \citet{iani_midis_2024}, {\it Virgil} is well described only when an AGN component is included, except for {\sc Synthesizer-AGN}, which tends to prefer a two-population solution. With respect to \citet{iani_midis_2024}, we expanded the two-population scenario also to {\sc Bagpipes}. 

For {\sc Bagpipes}, we adopt the same setup as in \citet{rinaldi_not_2024} for LRDs, utilizing synthetic templates from \citet{bruzual_stellar_2003} with a Kroupa IMF, a stellar mass cut-off of $100\,M_{\odot}$, and nebular emission modeled with \textsc{Cloudy} \citep{ferland_2013_2013}. We employ a continuity non-parametric SFH model \citep{leja_how_2019}, defining age bin edges (in look-back time) based on {\it Virgil}'s spectroscopic redshift, following a logarithmic distribution from $z=30$ to the cosmic time at that redshift. The same approach is applied to the age parameter.

We let stellar masses range from $10^5$ to $10^{13}\,M_{\odot}$ (uniform prior in log). We adopt a Calzetti reddening law \citep{calzetti_dust_2000}, allowing $A_{V}$ to vary between 0 and 6, while metallicity ($Z/Z_{\odot}$) spans 0 to 2.5. The ionization parameter is constrained between $-4$ and $0.001$. For the AGN contribution, we follow \citet{carnall_massive_2023} (see their Table 1), adopting broad, flat (uniform) priors for all parameters.

For the two-population scenario, we adopt a delayed SFH with $\tau$ ranging from 0.1 to 13 Gyr. Each population follows the same metallicity, age, and dust constraints as the AGN-included run.

For {\sc CIGALE}, we assume a delayed exponentially declining SFH (delayed-$\tau$ model) with two stellar populations. We adopt \citet{bruzual_stellar_2003} stellar population models with both solar and sub-solar metallicity ($Z = 0.2 Z_\odot$) and a Chabrier IMF. Nebular continuum and emission lines are included, assuming solar and sub-solar metallicity, and allowing electron density and ionization parameter values of $n_e = 10, 10^{2}, 10^{3}\,\rm cm^{-3}$ and $\log_{10}(\mathcal{U}) = -3, -2, -1$, respectively. For dust attenuation, we adopt \citet{calzetti_dust_2000}, while far-IR emission is modeled with \citet{draine_andromedas_2014}. AGN emission is incorporated using the SKIRTOR models \citep{stalevski_3d_2012, stalevski_dust_2016}, following initial parameters from \citet{yang_ceers_2023} but allowing for both Type 1 (unobscured) and Type 2 (obscured) AGN.

For {\sc Synthesizer-AGN}, the SED is modeled as a composite stellar population \citep{perez-gonzalez_stellar_2003, perez-gonzalez_stellar_2008} with AGN emission from the accretion disk and dusty torus \citep{perez-gonzalez_what_2024}. Stellar emission consists of a young and an older star formation event, each described by a delayed exponential function with $\tau$ between 1 Myr and 1 Gyr, and ages up to the cosmic time at the {\it Virgil}'s redshift. Each stellar population experiences independent attenuation following \citet{calzetti_dust_2000}, with $A_{V}$ ranging from 0 to 10 mag. The stellar component is modeled with \citet{bruzual_stellar_2003} templates, assuming a Chabrier IMF with stellar masses between 0.1 and $100\,M_\odot$, including nebular emission \citep{perez-gonzalez_stellar_2003}. The AGN emission follows a QSO average spectrum \citep{vanden_berk_composite_2001, glikman_near-infrared_2006}, while its dust component is modeled using self-consistent AGN torus templates from \citet{siebenmorgen_self-consistent_2015}.

Finally, for {\sc Prospector}, we used a modified version built on the standard implementation, which incorporates the Flexible Stellar Population Synthesis (FSPS; \citealt{conroy_fsps_2010}) model for the stellar component. We assumed a \citet{kroupa_variation_2001} initial mass function and a delayed-$\tau$ star formation history. Nebular line and continuum emission were included, as pre-configured in FSPS \citep{byler_nebular_2017}. We adopted the Calzetti attenuation curve with a flexible slope, following \citet{kriek_dust_2013}. For galaxy dust emission, given that {\it Virgil} has $z=6.64$, we used the empirical IR SED model of Haro~11, a low-metallicity, starbursting dwarf galaxy believed to exhibit typical features of first-generation galaxies in the early Universe \citep{lyu_contribution_2016, de_rossi_far-infrared_2018}. 

This code employs a set of semi-empirical AGN SED models optimized for AGN identification and characterization \citep{lyu_agn_2022, lyu_active_2024}. We adopted a model configuration similar to that in \citet{lyu_active_2024} for the SMILES+JADES AGN identification. The AGN component includes both the AGN-powered continuum from the UV to the far-IR and the narrow and broad emission lines from the UV to the NIR, derived from empirical observations. The continuum shape and line strengths of the AGN SED can be adjusted using a hybrid attenuation model, featuring an SMC-like curve for the typical UV-optical attenuation in Type-1 AGNs and an empirical attenuation law for IR obscuration. We refer the reader to \citet{lyu_active_2024} for further details.

\subsection{Results from SED fitting}

We now discuss the results of the SED fitting.
\citet{iani_midis_2024} showed that most SED fitting codes struggle to reproduce the MIRI fluxes unless an AGN component is included, with the exception of {\sc Synthesizer-AGN}. This is mainly because, by design, {\sc Synthesizer-AGN} always fits two independent stellar populations—regardless of the inclusion or not of an AGN—providing greater flexibility to account for complex SED shapes.

We first fit {\it Virgil}'s photometry without including an AGN component, adopting a model with two stellar populations. As shown in \citet{iani_midis_2024}, only {\sc Synthesizer-AGN} (used here without including an AGN component) was originally capable of such a fit due to its design. Here, we introduced this flexibility in {\sc Bagpipes} as well\footnote{This flexibility was implemented in {\sc Bagpipes} only for the model without the AGN contribution.}. In both cases, the inferred total $M_{\star}$ is significantly higher than in AGN-inclusive models (see next paragraph), with $\log_{10}(M_{\star}/M_{\odot})$ ranging from 9.80 to 10.50. The older population consistently exhibits $A_{V} \approx 2-4$. However, the solutions from {\sc Bagpipes} and {\sc Synthesizer-AGN} differ: the latter yields two very young populations ($t_{\mathrm{age}} \lesssim 50$ Myr) with distinct extinctions ($A_{V} \approx 3.60$ for the most obscured), consistent with \citet{iani_midis_2024}, while {\sc Bagpipes} retrieves two markedly distinct populations—one extremely young ($t_{\mathrm{age}} < 3$ Myr), dust-free, and low-mass ($\log_{10}(M_{\star}/M_{\odot}) \approx 7.24$), and one older (approaching the age of the Universe at $z \approx 6.64$), massive ($\log_{10}(M_{\star}/M_{\odot}) \approx 10.48$\footnote{Overlapping with the maximum $M_{\star}$ expected from the stellar mass–halo mass relation of \citet{behroozi_universe_2020} at its redshift.}), and heavily obscured ($A_{V} \approx 4$).


When including an AGN component, both {\sc Synthesizer-AGN} and {\sc Prospector} consistently yield $\log_{10}(M_{\star}/M_{\odot}) \approx 9$, although {\sc Bagpipes} retrieves a significantly lower stellar mass by $\approx 1$ dex: $\log_{10}(M_{\star}/M_{\odot}) = 7.92$. Similarly, {\sc CIGALE} finds $\log_{10}(M_{\star}/M_{\odot}) = 8.50$. Notably, {\sc Bagpipes} predicts a stellar age of $t_{\mathrm{age}} = 400_{-260}^{+260}$ Myr, consistent with both {\sc CIGALE} and {\sc Prospector} within the uncertainties ($t_{\mathrm{age}} \approx 400-600$ Myr). However, its mass-weighted age, from {\sc Bagpipes}, is only $4_{-3}^{+2}$ Myr, indicating that a substantial fraction of the stellar mass formed recently, which is consistent with a very recent burst of star formation.

Interestingly, both {\sc Synthesizer-AGN} and {\sc Prospector}, when invoking an AGN, predict $A_{V} \approx 2-4$. In particular, {\sc Synthesizer-AGN} predicts $A_{V} \approx 0.74$~mag for the young population ($t_{\mathrm{age}} \approx 10$ Myr), in line with our estimate from NIRSpec/PRISM data, and $A_{V} \approx 2.40$ for the older population ($t_{\mathrm{age}} \approx 32$ Myr). Similarly, {\sc Bagpipes} and {\sc CIGALE} yield $A_{V}$ values consistent with those inferred from Balmer lines tracing the young population. Across all fits, the metallicity remains consistent at $10$–$20\%, Z_{\odot}$, in agreement with our spectroscopic constraints.

\begin{figure*}
    \centering
    \includegraphics[width=1\linewidth]{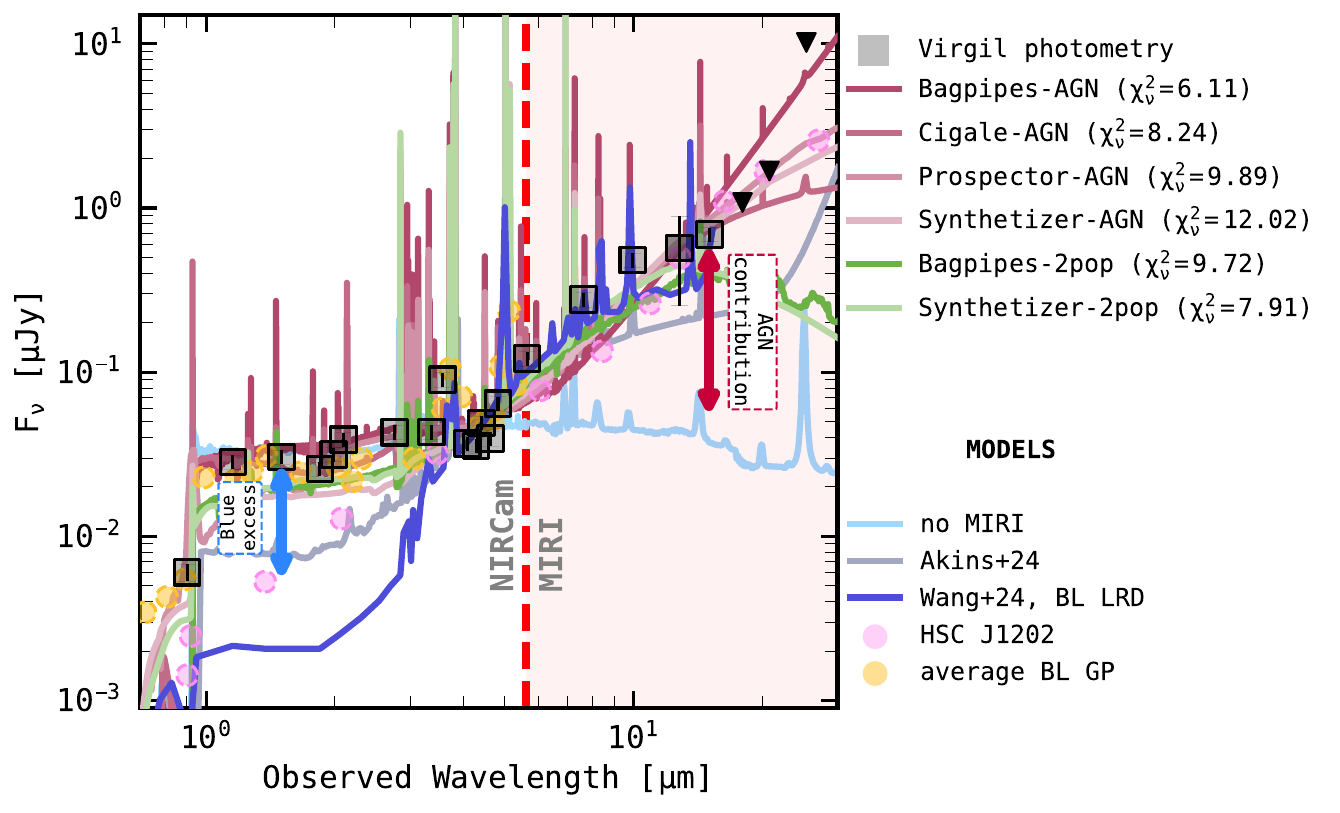}
    \caption{The best-fit results for {\it Virgil} are shown, with photometric data indicated by black squares and $3\sigma$ upper limits by black inverted triangles. The reduced chi-squared values are also reported and computed consistently across the different SED fitting codes to account for their varying treatments of upper limits. The red arrow highlights the IR excess relative to the typical SFG template, which both NIRCam and NIRSpec data would bias the SED fitting toward. In contrast, the blue arrow indicates the UV excess compared to the average LRD template proposed by \citet{akins_cosmos-web_2024}.      
    For comparison, we include the LRD model from \citet{akins_cosmos-web_2024} and {\it Virgil}'s SED, excluding the MIRI bands, as derived from {\sc bagpipes}. Additionally, we show the average BL GP SED (green circles) from \citet{lin_discovery_2024}, considering only sources with $M_{\bullet}$ similar to the estimated value for {\it Virgil}. Finally, we compare it with the Blue-excess HotDOG (HSC J1202; red circles) from \citet{noboriguchi_extreme_2022}. RUBIES-BLAGN-1 model from \citet{wang_rubies_2024} is shown in blue. The vertical dashed red line marks the transition between the NIRCam and MIRI domains. All models are normalized to F444W in the {\it Virgil}'s reference system.
}
    \label{fig:sed_fitting}
\end{figure*}

Overall, all models yield a reduced $\chi^{2}$ between 6 and 12, with {\sc Bagpipes} achieving the best $\chi^{2}_{\nu}$ when including an AGN, while {\sc Synthesizer-AGN} favors the model of two stellar populations without an AGN. As discussed in \citet{iani_midis_2024}, fitting this object remains challenging, reflecting the well-known difficulty in modeling LRDs in general. Nevertheless, most SED fitting codes used in this work—as it was for \citet{iani_midis_2024}—favor the (embedded) AGN interpretation. Interestingly, our modified version of {\sc Prospector} tailored for obscured AGNs \citep{lyu_active_2024} fails to reproduce the observed SED of this object when the AGN component is excluded. We show the best-fit results in Figure \ref{fig:sed_fitting} along with other models from the literature (see next section).

\section{Hide and Seek: What is the Nature of Virgil?}

\subsection{Star forming properties}

\citet{iani_midis_2024} introduced {\it Virgil} as the first LRD with a clearly detected host galaxy and a peculiar rising SED between 4 and 10~$\mu$m. At that time, their analysis was limited to photometric data from HST and JWST (NIRCam and MIRI). Leveraging multiple SED-fitting codes, we identified two plausible scenarios for {\it Virgil}'s nature: either a dusty starburst galaxy or a SFG hosting an obscured AGN. However, in the previous sections, we demonstrated that incorporating spectroscopic data from NIRSpec/PRISM strengthens the interpretation of \textit{Virgil} as a typical SFG during the EoR—or, at the very least, that its AGN nature becomes less clear when the redshift evolution of diagnostic diagrams is taken into account. This is supported by its relatively low $M_{\star}$, metal-poor nature, and extreme EW$_{0}$ for [O\,{\sc iii}]$\lambda\lambda4959,5007$. Interestingly, {\it Virgil} shares similarities with GPs and BBs, which are often considered the best low-redshift analogs of high-$z$ galaxies.

In general, when focusing on NIRSpec/PRISM data, we find that {\it Virgil} appears to be a typical system at $z\approx6-7$ undergoing a bursty star formation episode. At the time of observation, it is either experiencing or fading out of a starburst, while also harboring an older stellar population that dominates its $M_{\star}$, with formation dating back a few hundred million years, as inferred from {\sc bagpipes}, {\sc CIGALE}, and {\sc prospector} fits. {\it Virgil} exhibits extreme emission line ratios, including a high R3 index ($\approx 0.8$), consistent with other galaxies at similar redshifts \citep[e.g.,][]{cameron_jades_2023, tang_jwstnirspec_2023}. This is likely driven by intense star formation, with a surface SFR of $\Sigma_{SFR(\text{H}\alpha)}\approx5.25\,M_{\odot}\,\text{yr}^{-1}\,\text{kpc}^{-2}$ at a moderately low metallicity ($Z \approx 0.1-0.2\,Z_{\odot}$). Whether it contains an AGN remains ambiguous if the emission line ratios are interpreted in the context of its redshift and the evolution of these ratios,  consistent with trends observed in other LRDs by \citet{rinaldi_not_2024}. If this interpretation holds, it may also apply to other claimed Type 1 and Type 2 AGN in the literature \citep[e.g.,][]{kokorev_uncover_2023, scholtz_jades_2023}.

\subsection{{\it Virgil}'s AGN}

However, multiple SED-fitting codes that incorporate wavelengths beyond 5~$\mu$m suggest that this galaxy hosts a deeply obscured AGN, a characteristic commonly observed in many LRDs. Notably, {\it Virgil}'s infrared SED rises even more steeply than those of typical LRDs, making it one of the most extreme cases identified to date at $z \approx 6$–7. While {\it Virgil} may appear exceptional, it has lower-redshift analogs, such as the obscured AGNs identified in the HUDF by \citet{lyu_active_2024}. 

Interestingly, \citet{lin_discovery_2024} recently identified a subset of GP galaxies that exhibit the V-shaped SED commonly found in high-$z$ LRDs, along with broad emission lines, suggesting a possible evolutionary link between these two populations (see Figure~\ref{fig:sed_fitting}).  

Further exploring analogies with lower redshift sources provides key insights into the nature of {\it Virgil}. \citet{iani_midis_2024} measured a compactness of $c_{F444W} \approx 1.5$ for {\it Virgil}, consistent with the LRD criterion ($c_{F444W} < 2$; \citealt{kokorev_census_2024}). However, this threshold is somewhat arbitrary. If {\it Virgil} were at $z \approx 3.5$, for instance, the angular scale would be $\approx 7.3$~kpc/arcsec instead of 5.4~kpc/arcsec at $z \approx 6.64$, and the 4.44~$\mu$m emission would be even more dominated by the obscured AGN, which would make {\it Virgil} appear even more compact.

Another intriguing analogy is with the very red sources at $z \approx 2$ identified in {\it Spitzer} observations, commonly known as Dust Obscured Galaxies (DOGs) \citep[e.g.,][]{weedman_spitzer_2006, bussmann_star_2012}. The most extreme DOGs exhibit infrared SEDs that rise as steeply as that of {\it Virgil}, such as SST24 J142648.9+332927 \citep{bussmann_star_2012}. DOGs are typically classified into two subgroups based on their distinct SEDs: power law (PL) DOGs, indicative of AGN dominance, and “bump” DOGs, which show a flattening near rest-frame 1.6 $\mu$m, suggesting a significant contribution from a stellar population \citep[e.g.,][]{dey_significant_2008}. Therefore, “bump” DOGs are considered to correspond to galaxies in a star-forming phase (\citealt{, bussmann_hubble_2011}), while PL DOGs are associated with galaxies in an AGN phase (e.g., \citealt{bussmann_infrared_2009}).  \citet{dey_significant_2008} showed that the fraction of PL DOGs among all DOGs increases with increasing MIR flux density, similar to the luminosity dependence of the AGN fraction in ultraluminous infrared galaxies (ULIRGs; \citealt{sanders_luminous_1996, veilleux_optical_1999}). Notably, some of these features are commonly observed in LRDs as well \citep{perez-gonzalez_what_2024, williams_galaxies_2024}.

Hence, {\it Virgil} is likely experiencing a transition phase at the moment we observe it, allowing us to witness a starburst event with an embedded AGN. As shown in \citet{iani_midis_2024}, the MIRI part of {\it Virgil}'s SED cannot be solely attributed to star-formation heated dust emission. Otherwise, the required dust temperature would approach the typical sublimation temperature for the dust ($T_{\mathrm{subl}} \approx 1500-2000$ K, depending on dust composition; \citealt{temple_exploring_2021}), a scenario that would be difficult to reconcile with  “normal” star formation behavior \citep[e.g.,][]{sommovigo_warm_2020}. 

However, the DOG family is more diverse. For instance, HotDOGs are among the most luminous objects in the Universe, with bolometric luminosities often exceeding $10^{13-14} L_{\odot}$ (see an example of HotDOG, HSC J1202, in Figure \ref{fig:sed_fitting}). Discovered with the \textit{Wide-field Infrared Survey Explorer} \citep[WISE;][]{wright_wise_2010}, they exhibit extremely red MIR SEDs. Their emission is dominated by hyperluminous, highly obscured AGNs that power their intense mid-to-far infrared radiation (e.g., \citealt{eisenhardt_first_2012}). Interestingly, X-ray studies indicate that their obscuration levels are exceptionally high, often reaching or exceeding the Compton-thick limit ($N_H > 1.5 \times 10^{24}$ cm\(^2\)) (e.g., \citealt{ricci_growing_2017}). Due to this heavy obscuration, the host galaxy could dominate the UV and optical wavelengths, rather than direct AGN emission.

While most HotDOGs exhibit extreme dust obscuration that suppresses UV emission, a subset displays significant excess UV/optical light compared to the average population—these are known as Blue-excess HotDOGs (BHDs; e.g., \citealt{assef_hot_2016, noboriguchi_extreme_2022}). \citet{assef_hot_2016} proposed several plausible scenarios to explain this excess, including the possibility that a small fraction of AGN light leaks into our line of sight via scattering by dust, gas, or both—a scenario recently studied also by \citet{stepney_big_2024} in the case of ULASJ2315+0143, a source around Cosmic Noon. In this scenario, the intrinsic emission from the accretion disk and broad-line region is largely absorbed, with only $\lesssim1\%$ scattered into view on larger spatial scales, where dust grains efficiently reflect shorter-wavelength radiation \citep{draine_scattering_2003}. 


A similar contrast between the UV and optical colors has been already observed in LRDs (\citealt{noboriguchi_similarity_2023}).  While most LRDs exhibit extremely red optical-to-NIR colors ($0.4$–$1~\mu$m), they show a blue excess in the rest-frame UV ($0.2$–$0.4~\mu$m), reminiscent of Blue-Excess HotDOGs found at $z\approx2-3$. However, LRDs represent a less extreme manifestation of this phenomenon compared to Blue-Excess HotDOGs, as they are found at higher redshifts and exhibit lower $M_{\bullet}$ and Eddington ratios ($\lambda_{\rm Edd} \approx 0.1$–$1$), suggesting they may be earlier-stage counterparts.

Nonetheless, {\it Virgil} stands out as an even more extreme case among LRDs, exhibiting both an exceptionally steep SED between F444W and F1500W and an enhanced UV excess. Indeed, {\it Virgil}’s UV excess is more pronounced compared to the average LRD template proposed by \citet{akins_cosmos-web_2024}, suggesting a greater contribution from either scattered AGN light or, more likely, a strong star-forming component. At the same time, its SED rises more steeply than seen in most LRDs at wavelengths beyond 0.6 $\mu$m (rest-frame), resembling the most obscured DOGs. This dual nature—both an extreme red excess in the MIRI regime and an unusually strong blue excess in the UV—potentially places Virgil as a high-$z$ analog of Blue-excess HotDOGs, possibly with a strong contribution at the UV regime from its host.

\citet{noboriguchi_similarity_2023} suggested that the connection between LRDs and Blue-excess HotDOGs reflects an evolutionary sequence in which gas-rich mergers trigger dusty starbursts, giving rise to heavily obscured DOGs. Interestingly, when modeling {\it Virgil} with {\sc forcepho}, it appears to be composed of two distinct knots in the NIRCam wavelength range, suggesting a possible merger-driven scenario. As AGN feedback disperses the surrounding dust, some of these systems develop a blue excess, akin to what is observed in Blue-excess HotDOGs. In this context, LRDs may represent an even earlier stage in this process, where lower metallicities and dust-to-gas ratios at high redshifts lead to a higher fraction of such objects exhibiting blue excess. Yet, {\it Virgil} appears even more extreme, with a UV excess likely arising from a combination of AGN activity and star formation, though primarily driven by the host galaxy, while its obscured SMBH shapes an extreme near-IR SED (rest-frame).

As a final analogy with sources at lower redshifts, in Figure~\ref{fig:sed_fitting}, we compared {\it Virgil} to another LRD at a much lower redshift ($z_{\text{spec}} = 3.1$), RUBIES-BLAGN-1, studied by \citet{wang_rubies_2024}. RUBIES-BLAGN-1 closely follows {\it Virgil}'s SED at wavelengths beyond 4.4 $\mu$m. While both objects exhibit a similar trend at IR wavelengths, they diverge significantly in the UV, further emphasizing the striking UV excess observed in {\it Virgil} compared to the average LRD populaton. This divergence highlights once again the peculiar nature of {\it Virgil}.

\section{Summary and Conclusions}

In this paper, we extend the analysis of {\it Virgil}, first presented in \citet{iani_midis_2024}, using recent data collected by the MIDIS and PAHSPECS teams for MIRI, and NIRSpec/PRISM data by the OASIS team. {\it Virgil} is a LAE at $z_{spec} = 6.6379 \pm 0.003$, detected with VLT/MUSE \citep{bacon_muse_2023}. It shares key photometric properties with the LRD population, including its compactness in F444W ($c_{F444W}\approx1.5$; \citealt{iani_midis_2024}), although it does not strictly meet all their color selection criteria (\citealt{kokorev_census_2024, rinaldi_not_2024}).

Our main findings can be summarized as follows:

\begin{itemize}
\item \textit{Virgil} exhibits low to moderate dust attenuation based on its Balmer decrement, assuming the SMC reddening law from \citet{gordon_quantitative_2003}. Its metallicity aligns with galaxies of similar $M_{\star}$ at $z \gtrsim 6$ (Figure \ref{fig:gas_phase_metallicity}). By comparing the UV- and H$\alpha$-based SFRs, we find that \textit{Virgil} may be transitioning into—or fading out of—a bursty phase. Its $f_{esc,LyC}$ and $\xi_{ion}$ indicate a limited role in Cosmic Reionization, further supported by its moderate EW$_{0}$(H$\alpha$) compared to other emitters at similar redshifts \citep[e.g.,][]{rinaldi_midis_2023, rinaldi_midis_2024, simmonds_low-mass_2024}. Overall, {\it Virgil}'s spectral properties align with the average galaxy population during the EoR (e.g., \citealt{cameron_jades_2023}). A full summary of its spectral properties is provided in Table \ref{tab:galaxy_em_prop}.

\item The combination of NIRSpec/PRISM data with extensive photometric coverage from HST and JWST, including MIRI detections up to 15~$\mu$m, offers a unique opportunity to further investigate {\it Virgil}'s nature. Without an {\it a priori} assumption that it could host an AGN— as previously suggested by \citet{iani_midis_2024} and based solely on photometry—we analyze its ISM properties using the R23 vs. O32 and Ne3O2Hd vs. Ne3O2 diagrams (Figure \ref{fig:ism}). By considering {\sc CLOUDY} models for SFGs (following the approach outlined in \citealt{calabro_evidence_2024}), we find that {\it Virgil}'s ISM does not significantly deviate from the typical high-$z$ SFGs during the EoR. Interestingly, its properties resemble well-studied GPs and BBs, which are among the best local analogs of high-$z$ galaxies. Notably, despite its moderate EW$_{0}$(H$\alpha$), {\it Virgil} exhibits a huge EW$_{0}$ for [O\,{\sc iii}]$\lambda\lambda4959,5007$ ($1514\pm20$~\AA), comparable to the Extreme Line Emitters studied by \citet{boyett_extreme_2024} at similar redshifts.

\item To assess the AGN nature suggested by \citet{iani_midis_2024}, we search for broad Balmer lines (a feature already found in other claimed LRDs), modeling H$\alpha$ with and without a broad component. However, the NIRSpec/PRISM data do not provide a firm conclusion (Figure~\ref{fig:broad_no_broad_Ha}). We note that the source's position within the slit (Figure~\ref{fig:virgil_spectrum}, right panel)—with the red knot lying near the edge or possibly in the shutter gap—may limit our sensitivity to broad-line emission. Nonetheless, assuming the broadening is real, the inferred $M_{\bullet}$, $L_{Bol}$, and Eddington ratio align with expectations for LRDs (Figure~\ref{fig:black_hole_prop}) and with values already reported in \citet{iani_midis_2024} by leveraging photometric data only. We examine emission-line diagnostics and find that while {\it Virgil} lies above the SFG-AGN separation line in the “OHNO” diagram \citep{backhaus_clear_2022}, it falls below it when accounting for redshift evolution. Similarly, O3Hg-based diagnostics \citep{mazzolari_new_2024} place it among AGNs at high $z$ \citep[e.g.,][]{nakajima_jwst_2023, scholtz_jades_2023}, but when accounting for redshift evolution, its classification becomes ambiguous (Figure~\ref{fig:virgil_line_ratios_oiii_hg}). Overall, while these high-$z$ diagnostics would classify {\it Virgil} as an AGN, they do not provide a definitive classification when the redshift evolution is taken into account, instead pointing to a likely mixed nature.

\item The newly acquired MIRI data from MIDIS and PAHSPECS, including a clear 15~$\mu$m detection of {\it Virgil}, enabled improved photometry and SED fitting. Using {\sc ForcePho}, we identified two distinct knots, reinforcing evidence of complex UV morphology in LRDs \citep{rinaldi_not_2024}, with one component accounting for the MIRI excess. {\it Virgil} appears even redder in F444W$-$F1500W color ($\approx2.84$) than typical LRDs and, while it does not meet all standard LRD color criteria (\citealt{kokorev_census_2024}), it satisfies the V-shaped SED selection from \citet{kocevski_rise_2024}, suggesting that MIRI bands may be crucial for identifying extreme cases (Figure~\ref{fig:virgil_color_mag}). To contextualize its spectral properties, we applied multiple SED-fitting codes (Figure~\ref{fig:sed_fitting}), extending the analysis of \citet{iani_midis_2024} by incorporating two independent stellar populations in {\sc bagpipes} and utilizing an AGN-optimized version of {\sc Prospector} \citep{lyu_active_2024}.  Interestingly, the tailored version of {\sc Prospector} fails to reproduce the observed SED when an AGN component is not included. While modeling this source remains challenging, our results show that the fits consistently require the presence of a dust-obscured AGN, consistent with the findings of \citet{iani_midis_2024}.

\item Combining photometric and spectroscopic data, we find {\it Virgil} to be among the most extreme LRDs identified to date, with a steeply rising SED beyond 4.44 $\mu$m (observed-frame) and a pronounced UV excess. Its SED closely resembles Cosmic Noon objects observed with {\it Spitzer}, such as SST24 J142648.9+33292, a well-studied DOG, and shares similarities with HotDOGs, whose IR properties align with those of LRDs. \citet{noboriguchi_similarity_2023} linked JWST/EROs to DOGs, particularly Blue-Excess HotDOGs, which exhibit extreme IR colors and a flat UV continuum. The UV excess in Blue-Excess HotDOGs has been attributed to AGN light ($<1\%$; \citealt{assef_hot_2016, assef_hot_2020}) scattered by dust or gas, a scenario that could also contribute to {\it Virgil}'s properties. A similar finding was also reported for  ULASJ2315+014, a Cosmic Noon source, by \citet{stepney_big_2024}. However, {\it Virigil}
s SED and emission-line features suggest a mix of AGN and star formation activity, with the UV emission likely dominated by the host galaxy. Finally, we compare Virgil to RUBIES-BLAGN-1 ($z_{spec} = 3.1$) from \citet{wang_rubies_2024}, noting that their SEDs diverge at shorter wavelengths, with {\it Virgil} exhibiting a stronger UV excess.
\end{itemize}

This study underscores the challenges of studying such systems at high redshift and highlights the necessity of a multi-wavelength approach to identify and characterize extreme sources during Cosmic Reionization. Deep MIRI imaging at $z > 6$ is crucial for uncovering highly dust-obscured AGNs that might otherwise be misclassified as normal SFGs, especially in studies limited to wavelengths below 5~$\mu$m. In such cases, both photometric and spectroscopic information can lead to ambiguous classifications, particularly when redshift evolution is taken into account. This reinforces the need to expand MIRI coverage to systematically identify more of these extreme objects and refine our understanding of early black hole growth.

\acknowledgments
The authors deeply thank Antonello Calabrò for valuable insights on {\sc CLOUDY} and {\sc pyCloudy} and for publicly sharing their SFG and AGN models. They also express their gratitude to Adam Carnall for their inputs on {\sc bagpipes} and to Camilla Pacifici, Vasily Kokorev, and Cristian Vignali for their insightful discussions.

\smallskip

This work is based on observations made with the NASA/ESA/CSA JWST. The data were obtained from the Mikulski Archive for Space Telescopes (MAST) at the Space Telescope Science Institute, which is operated by the Association of Universities for Research in Astronomy, Inc., under NASA contract NAS 5-03127 for JWST. These observations are associated with JWST programs GTO \#1180, GO \#1210,  GTO\#1283, GO \#1963, GO \#1895, and GO\# 3215, and GO\#6511.

The authors acknowledge the FRESCO, JEMS, and \# 3215 teams led by coPIs P. Oesch, C. C. Williams, M. Maseda, D. Eisenstein, and R. Maiolino for developing their observing program with a zero-exclusive-access period. Processing for the JADES NIRCam data release was performed on the lux cluster at the University of California, Santa Cruz, funded by NSF MRI grant AST 1828315. Also based on observations made with the NASA/ESA Hubble Space Telescope obtained from the Space Telescope Science Institute, which is operated by the Association of Universities for Research in Astronomy, Inc., under NASA contract NAS 526555. The data presented in this article were obtained from MAST at the Space Telescope Science Institute. The specific observations analyzed can be accessed via \dataset[DOI: 10.17909/gdyc-7g80, 10.17909/fsc4-dt61, 10.17909/fsc4-dt61, 10.17909/T91019, 10.17909/1rq3-8048, 10.17909/z2gw-mk31]..

\smallskip

The authors acknowledge use of the lux supercomputer at UC Santa Cruz, funded by NSF MRI grant AST 1828315.

AJB acknowledge funding from the "FirstGalaxies" Advanced Grant from the European Research Council (ERC) under the European Union’s Horizon 2020 research and innovation programme (Grant agreement No. 789056)

PGP-G acknowledges support from grant PID2022-139567NB-I00 funded by Spanish Ministerio de Ciencia e Innovaci\'on MCIN/AEI/10.13039/501100011033, FEDER, UE.

BER acknowledges support from the NIRCam Science Team contract to the University of Arizona, NAS5-02015, and JWST Program 3215.

ST acknowledges support by the Royal Society Research Grant G125142.

The research of CCW is supported by NOIRLab, which is managed by the Association of Universities for Research in Astronomy (AURA) under a cooperative agreement with the National Science Foundation.

JW gratefully acknowledges support from the Cosmic Dawn Center through the DAWN Fellowship. The Cosmic Dawn Center (DAWN) is funded by the Danish National Research Foundation under grant No. 140.

YZ, ZJ gratefully acknowledges JWST/NIRCam contract to the University of Arizona NAS5-02015.

H\"U acknowledges funding by the European Union (ERC APEX, 101164796). Views and opinions expressed are however those of the authors only and do not necessarily reflect those of the European Union or the European Research Council Executive Agency. Neither the European Union nor the granting authority can be held responsible for them.

GCJ acknowledges support by the Science and Technology Facilities Council (STFC), ERC Advanced Grant 695671 "QUENCH".

ACG acknowledges support by JWST contract B0215/JWST-GO-02926.

GO acknowledges support from the Swedish National Space Agency (SNSA)

The Cosmic Dawn Center (DAWN) is funded by the Danish National Research Foundation (DNRF) under grant No. 140.

HI acknowledges support from JSPS KAKENHI grant No. JP21H01129.

MA gratefully acknowledges support from ANID Basal Project FB210003 and ANID MILENIO NCN2024\_112.

TDS acknowledges the research project was supported by the Hellenic Foundation for Research and Innovation (HFRI) under the “2nd Call for HFRI Research Projects to support Faculty Members \& Researchers” (Project Number: 03382)

RM acknowledges support by the Science and Technology Facilities Council (STFC), by the ERC through Advanced Grant 695671 “QUENCH”, and by the UKRI Frontier Research grant RISEandFALL. RM also acknowledges funding from a research professorship from the Royal Society.

IS acknowledges funding Atracc{\' i}on de Talento Grant No.2022-T1/TIC-20472 of the Comunidad de Madrid, Spain and the European Research Council (ERC) under the European Union’s Horizon 2020 research and innovation programme (Grant No. 101117541, DistantDust).

KIC acknowledges funding from the Dutch Research Council (NWO) through
the award of the Vici Grant VI.C.212.036.

\vspace{5mm}
\facilities{{\sl HST}, {\sl JWST}}.

\software{\textsc{Astropy} \citep{astropy_collaboration_astropy_2022}, 
\textsc{Bagpipes}
\citep{carnall_vandels_2019},
\textsc{MSAEXP}
\citep{brammer_msaexp_2023}
          \textsc{NumPy} \citep{harris_array_2020},
          \textsc{pandas} \citep{team_pandas-devpandas_2024}
          \textsc{Photutils} \citep{bradley_photutils_2016}, 
          \textsc{TOPCAT} \citep{taylor_topcat_2022}.
          }
\bibliography{references}{}
\bibliographystyle{aasjournal}

\end{document}